\documentclass[longauth]{aa}

\usepackage{graphicx}
\usepackage{natbib}
\usepackage{scalerel}
\usepackage{comment}

\usepackage[table]{xcolor}

\usepackage{txfonts}
\usepackage[pdfencoding=auto,psdextra]{hyperref}
\hypersetup{
    colorlinks=true,
    linkcolor=blue,
    filecolor=magenta,      
    urlcolor=blue,
    citecolor=blue
}
\makeatletter
\renewcommand*\aa@pageof{, page \thepage{} of \pageref*{LastPage}}
\makeatother

\usepackage[utf8]{inputenc}

\usepackage[switch, modulo]{lineno}
              
\renewcommand{\linenumbers}[0]{}

\usepackage{euclid}

\begin{document}

\title{\Euclid: Calibrating distances from surface brightness fluctuations with Early Release Observations of the Fornax cluster\thanks{This paper is published on behalf of the Euclid Consortium.}}    

\newcommand{\orcid}[1]{} 			   
 %%%% Version Thursday 25th of June 2026 12:52:06 PM UT
%%%% Assumes the new A&A style file from Oct 2025 or later
%%%% Please do not edit the author list -- contact ECEB Bureau for changes
%\newcommand{\orcid}[1]{} %% if already defined in aa.cls: comment, or use renewcommand			   
\author{R.~Habas\orcid{0000-0002-4033-3841}\thanks{\email{rebecca.habas@inaf.it}}\inst{\ref{aff1}}
\and M.~Cantiello\orcid{0000-0003-2072-384X}\inst{\ref{aff1}}
\and G.~Riccio\orcid{0000-0002-6399-2129}\inst{\ref{aff1}}
\and G.~Raimondo\orcid{0000-0002-5577-7023}\inst{\ref{aff1}}
\and N.~Hazra\orcid{0000-0002-3870-1537}\inst{\ref{aff2},\ref{aff3},\ref{aff1}}
\and S.~Mei\orcid{0000-0002-2849-559X}\inst{\ref{aff4},\ref{aff5}}
\and A.~Lan\c{c}on\orcid{0000-0002-7214-8296}\inst{\ref{aff6}}
\and G.~D'Ago\orcid{0000-0001-9697-7331}\inst{\ref{aff7}}
\and R.~Scaramella\orcid{0000-0003-2229-193X}\inst{\ref{aff8}}
\and L.~K.~Hunt\orcid{0000-0001-9162-2371}\inst{\ref{aff9}}
\and J.-C.~Cuillandre\orcid{0000-0002-3263-8645}\inst{\ref{aff10}}
\and J.~P.~Blakeslee\orcid{0000-0002-5213-3548}\inst{\ref{aff11},\ref{aff12}}
\and J.~B.~Jensen\orcid{0000-0001-8762-8906}\inst{\ref{aff13}}
\and E.~Brocato\orcid{0000-0001-7988-8177}\inst{\ref{aff8}}
\and D.~Carollo\orcid{0000-0002-0005-5787}\inst{\ref{aff14}}
\and P.-A.~Duc\orcid{0000-0003-3343-6284}\inst{\ref{aff6}}
\and M.~Kluge\orcid{0000-0002-9618-2552}\inst{\ref{aff15}}
\and R.~Laureijs\inst{\ref{aff16}}
\and O.~Marchal\orcid{0000-0001-7461-8928}\inst{\ref{aff6}}
\and F.~R.~Marleau\orcid{0000-0002-1442-2947}\inst{\ref{aff17}}
\and M.~Mirabile\orcid{0009-0007-6055-3933}\inst{\ref{aff1},\ref{aff3}}
\and R.~F.~Peletier\orcid{0000-0001-7621-947X}\inst{\ref{aff16}}
\and M.~Poulain\orcid{0000-0002-7664-4510}\inst{\ref{aff18}}
\and R.~S\'anchez-Janssen\orcid{0000-0003-4945-0056}\inst{\ref{aff19}}
\and T.~Saifollahi\orcid{0000-0002-9554-7660}\inst{\ref{aff6}}
\and M.~Schirmer\orcid{0000-0003-2568-9994}\inst{\ref{aff20}}
\and E.~Sola\orcid{0000-0002-2814-3578}\inst{\ref{aff7}}
\and M.~Urbano\orcid{0000-0001-5640-0650}\inst{\ref{aff21},\ref{aff6}}
\and K.~Voggel\orcid{0000-0001-6215-0950}\inst{\ref{aff6}}
\and B.~Altieri\orcid{0000-0003-3936-0284}\inst{\ref{aff22}}
\and S.~Andreon\orcid{0000-0002-2041-8784}\inst{\ref{aff23}}
\and N.~Auricchio\orcid{0000-0003-4444-8651}\inst{\ref{aff24}}
\and C.~Baccigalupi\orcid{0000-0002-8211-1630}\inst{\ref{aff25},\ref{aff14},\ref{aff26},\ref{aff27}}
\and M.~Baldi\orcid{0000-0003-4145-1943}\inst{\ref{aff28},\ref{aff24},\ref{aff29}}
\and A.~Balestra\orcid{0000-0002-6967-261X}\inst{\ref{aff30}}
\and S.~Bardelli\orcid{0000-0002-8900-0298}\inst{\ref{aff24}}
\and P.~Battaglia\orcid{0000-0002-7337-5909}\inst{\ref{aff24}}
\and R.~Bender\orcid{0000-0001-7179-0626}\inst{\ref{aff15},\ref{aff31}}
\and A.~Biviano\orcid{0000-0002-0857-0732}\inst{\ref{aff14},\ref{aff25}}
\and E.~Branchini\orcid{0000-0002-0808-6908}\inst{\ref{aff32},\ref{aff33},\ref{aff23}}
\and M.~Brescia\orcid{0000-0001-9506-5680}\inst{\ref{aff34},\ref{aff35}}
\and S.~Camera\orcid{0000-0003-3399-3574}\inst{\ref{aff36},\ref{aff37},\ref{aff38}}
\and V.~Capobianco\orcid{0000-0002-3309-7692}\inst{\ref{aff38}}
\and C.~Carbone\orcid{0000-0003-0125-3563}\inst{\ref{aff39}}
\and V.~F.~Cardone\inst{\ref{aff8},\ref{aff40}}
\and J.~Carretero\orcid{0000-0002-3130-0204}\inst{\ref{aff41},\ref{aff42}}
\and S.~Casas\orcid{0000-0002-4751-5138}\inst{\ref{aff43},\ref{aff44}}
\and M.~Castellano\orcid{0000-0001-9875-8263}\inst{\ref{aff8}}
\and G.~Castignani\orcid{0000-0001-6831-0687}\inst{\ref{aff24}}
\and S.~Cavuoti\orcid{0000-0002-3787-4196}\inst{\ref{aff35},\ref{aff45}}
\and K.~C.~Chambers\orcid{0000-0001-6965-7789}\inst{\ref{aff46}}
\and A.~Cimatti\inst{\ref{aff47}}
\and C.~Colodro-Conde\inst{\ref{aff48}}
\and G.~Congedo\orcid{0000-0003-2508-0046}\inst{\ref{aff49}}
\and C.~J.~Conselice\orcid{0000-0003-1949-7638}\inst{\ref{aff50}}
\and L.~Conversi\orcid{0000-0002-6710-8476}\inst{\ref{aff51},\ref{aff22}}
\and Y.~Copin\orcid{0000-0002-5317-7518}\inst{\ref{aff52}}
\and F.~Courbin\orcid{0000-0003-0758-6510}\inst{\ref{aff53},\ref{aff54},\ref{aff55}}
\and H.~M.~Courtois\orcid{0000-0003-0509-1776}\inst{\ref{aff56}}
\and M.~Cropper\orcid{0000-0003-4571-9468}\inst{\ref{aff57}}
\and H.~Degaudenzi\orcid{0000-0002-5887-6799}\inst{\ref{aff58}}
\and G.~De~Lucia\orcid{0000-0002-6220-9104}\inst{\ref{aff14}}
\and H.~Dole\orcid{0000-0002-9767-3839}\inst{\ref{aff59}}
\and F.~Dubath\orcid{0000-0002-6533-2810}\inst{\ref{aff58}}
\and X.~Dupac\inst{\ref{aff22}}
\and S.~Escoffier\orcid{0000-0002-2847-7498}\inst{\ref{aff60}}
\and M.~Fabricius\orcid{0000-0002-7025-6058}\inst{\ref{aff15},\ref{aff31}}
\and M.~Farina\orcid{0000-0002-3089-7846}\inst{\ref{aff61}}
\and R.~Farinelli\inst{\ref{aff24}}
\and S.~Ferriol\inst{\ref{aff52}}
\and S.~Fotopoulou\orcid{0000-0002-9686-254X}\inst{\ref{aff62}}
\and M.~Frailis\orcid{0000-0002-7400-2135}\inst{\ref{aff14}}
\and E.~Franceschi\orcid{0000-0002-0585-6591}\inst{\ref{aff24}}
\and M.~Fumana\orcid{0000-0001-6787-5950}\inst{\ref{aff39}}
\and S.~Galeotta\orcid{0000-0002-3748-5115}\inst{\ref{aff14}}
\and K.~George\orcid{0000-0002-1734-8455}\inst{\ref{aff63}}
\and B.~Gillis\orcid{0000-0002-4478-1270}\inst{\ref{aff49}}
\and C.~Giocoli\orcid{0000-0002-9590-7961}\inst{\ref{aff24},\ref{aff29}}
\and P.~G\'omez-Alvarez\orcid{0000-0002-8594-5358}\inst{\ref{aff64},\ref{aff22}}
\and J.~Gracia-Carpio\orcid{0000-0003-4689-3134}\inst{\ref{aff15}}
\and A.~Grazian\orcid{0000-0002-5688-0663}\inst{\ref{aff30}}
\and F.~Grupp\inst{\ref{aff15},\ref{aff31}}
\and S.~V.~H.~Haugan\orcid{0000-0001-9648-7260}\inst{\ref{aff65}}
\and H.~Hoekstra\orcid{0000-0002-0641-3231}\inst{\ref{aff66}}
\and W.~Holmes\orcid{0009-0007-8554-4646}\inst{\ref{aff67}}
\and I.~M.~Hook\orcid{0000-0002-2960-978X}\inst{\ref{aff68}}
\and F.~Hormuth\inst{\ref{aff69}}
\and A.~Hornstrup\orcid{0000-0002-3363-0936}\inst{\ref{aff70},\ref{aff71}}
\and K.~Jahnke\orcid{0000-0003-3804-2137}\inst{\ref{aff20}}
\and M.~Jhabvala\inst{\ref{aff72}}
\and B.~Joachimi\orcid{0000-0001-7494-1303}\inst{\ref{aff73}}
\and S.~Kermiche\orcid{0000-0002-0302-5735}\inst{\ref{aff60}}
\and A.~Kiessling\orcid{0000-0002-2590-1273}\inst{\ref{aff67}}
\and B.~Kubik\orcid{0009-0006-5823-4880}\inst{\ref{aff52}}
\and M.~K\"ummel\orcid{0000-0003-2791-2117}\inst{\ref{aff31}}
\and M.~Kunz\orcid{0000-0002-3052-7394}\inst{\ref{aff74}}
\and H.~Kurki-Suonio\orcid{0000-0002-4618-3063}\inst{\ref{aff75},\ref{aff76}}
\and A.~M.~C.~Le~Brun\orcid{0000-0002-0936-4594}\inst{\ref{aff77}}
\and P.~B.~Lilje\orcid{0000-0003-4324-7794}\inst{\ref{aff65}}
\and V.~Lindholm\orcid{0000-0003-2317-5471}\inst{\ref{aff75},\ref{aff76}}
\and I.~Lloro\orcid{0000-0001-5966-1434}\inst{\ref{aff78}}
\and G.~Mainetti\orcid{0000-0003-2384-2377}\inst{\ref{aff79}}
\and O.~Mansutti\orcid{0000-0001-5758-4658}\inst{\ref{aff14}}
\and O.~Marggraf\orcid{0000-0001-7242-3852}\inst{\ref{aff80}}
\and M.~Martinelli\orcid{0000-0002-6943-7732}\inst{\ref{aff8},\ref{aff40}}
\and N.~Martinet\orcid{0000-0003-2786-7790}\inst{\ref{aff81}}
\and F.~Marulli\orcid{0000-0002-8850-0303}\inst{\ref{aff82},\ref{aff24},\ref{aff29}}
\and R.~J.~Massey\orcid{0000-0002-6085-3780}\inst{\ref{aff83}}
\and E.~Medinaceli\orcid{0000-0002-4040-7783}\inst{\ref{aff24}}
\and M.~Melchior\inst{\ref{aff84}}
\and M.~Meneghetti\orcid{0000-0003-1225-7084}\inst{\ref{aff24},\ref{aff29}}
\and E.~Merlin\orcid{0000-0001-6870-8900}\inst{\ref{aff8}}
\and G.~Meylan\inst{\ref{aff85}}
\and A.~Mora\orcid{0000-0002-1922-8529}\inst{\ref{aff86}}
\and M.~Moresco\orcid{0000-0002-7616-7136}\inst{\ref{aff82},\ref{aff24}}
\and L.~Moscardini\orcid{0000-0002-3473-6716}\inst{\ref{aff82},\ref{aff24},\ref{aff29}}
\and R.~Nakajima\orcid{0009-0009-1213-7040}\inst{\ref{aff80}}
\and C.~Neissner\orcid{0000-0001-8524-4968}\inst{\ref{aff87},\ref{aff42}}
\and R.~C.~Nichol\orcid{0000-0003-0939-6518}\inst{\ref{aff88}}
\and S.-M.~Niemi\orcid{0009-0005-0247-0086}\inst{\ref{aff89}}
\and J.~W.~Nightingale\orcid{0000-0002-8987-7401}\inst{\ref{aff90}}
\and C.~Padilla\orcid{0000-0001-7951-0166}\inst{\ref{aff87}}
\and S.~Paltani\orcid{0000-0002-8108-9179}\inst{\ref{aff58}}
\and F.~Pasian\orcid{0000-0002-4869-3227}\inst{\ref{aff14}}
\and K.~Pedersen\inst{\ref{aff91}}
\and W.~J.~Percival\orcid{0000-0002-0644-5727}\inst{\ref{aff92},\ref{aff93},\ref{aff94}}
\and V.~Pettorino\orcid{0000-0002-4203-9320}\inst{\ref{aff89}}
\and A.~Pezzotta\orcid{0000-0003-0726-2268}\inst{\ref{aff23}}
\and S.~Pires\orcid{0000-0002-0249-2104}\inst{\ref{aff10}}
\and G.~Polenta\orcid{0000-0003-4067-9196}\inst{\ref{aff95}}
\and M.~Poncet\inst{\ref{aff96}}
\and L.~A.~Popa\inst{\ref{aff97}}
\and L.~Pozzetti\orcid{0000-0001-7085-0412}\inst{\ref{aff24}}
\and F.~Raison\orcid{0000-0002-7819-6918}\inst{\ref{aff15}}
\and R.~Rebolo\orcid{0000-0003-3767-7085}\inst{\ref{aff48},\ref{aff98},\ref{aff99}}
\and A.~Renzi\orcid{0000-0001-9856-1970}\inst{\ref{aff100},\ref{aff101},\ref{aff24}}
\and J.~Rhodes\orcid{0000-0002-4485-8549}\inst{\ref{aff67}}
\and G.~Riccio\inst{\ref{aff35}}
\and E.~Romelli\orcid{0000-0003-3069-9222}\inst{\ref{aff14}}
\and M.~Roncarelli\orcid{0000-0001-9587-7822}\inst{\ref{aff24}}
\and B.~Rusholme\orcid{0000-0001-7648-4142}\inst{\ref{aff102}}
\and R.~Saglia\orcid{0000-0003-0378-7032}\inst{\ref{aff31},\ref{aff15}}
\and Z.~Sakr\orcid{0000-0002-4823-3757}\inst{\ref{aff103},\ref{aff104},\ref{aff105}}
\and D.~Sapone\orcid{0000-0001-7089-4503}\inst{\ref{aff106}}
\and B.~Sartoris\orcid{0000-0003-1337-5269}\inst{\ref{aff31},\ref{aff14}}
\and P.~Schneider\orcid{0000-0001-8561-2679}\inst{\ref{aff80}}
\and M.~Scodeggio\inst{\ref{aff39}}
\and A.~Secroun\orcid{0000-0003-0505-3710}\inst{\ref{aff60}}
\and E.~Sihvola\orcid{0000-0003-1804-7715}\inst{\ref{aff107}}
\and P.~Simon\inst{\ref{aff80}}
\and C.~Sirignano\orcid{0000-0002-0995-7146}\inst{\ref{aff100},\ref{aff101}}
\and G.~Sirri\orcid{0000-0003-2626-2853}\inst{\ref{aff29}}
\and L.~Stanco\orcid{0000-0002-9706-5104}\inst{\ref{aff101}}
\and P.~Tallada-Cresp\'{i}\orcid{0000-0002-1336-8328}\inst{\ref{aff41},\ref{aff42}}
\and A.~N.~Taylor\inst{\ref{aff49}}
\and H.~I.~Teplitz\orcid{0000-0002-7064-5424}\inst{\ref{aff108}}
\and I.~Tereno\orcid{0000-0002-4537-6218}\inst{\ref{aff109},\ref{aff110}}
\and N.~Tessore\orcid{0000-0002-9696-7931}\inst{\ref{aff57}}
\and S.~Toft\orcid{0000-0003-3631-7176}\inst{\ref{aff111},\ref{aff112}}
\and R.~Toledo-Moreo\orcid{0000-0002-2997-4859}\inst{\ref{aff113}}
\and F.~Torradeflot\orcid{0000-0003-1160-1517}\inst{\ref{aff42},\ref{aff41}}
\and I.~Tutusaus\orcid{0000-0002-3199-0399}\inst{\ref{aff114},\ref{aff115},\ref{aff104}}
\and J.~Valiviita\orcid{0000-0001-6225-3693}\inst{\ref{aff75},\ref{aff76}}
\and T.~Vassallo\orcid{0000-0001-6512-6358}\inst{\ref{aff14},\ref{aff63}}
\and G.~Verdoes~Kleijn\orcid{0000-0001-5803-2580}\inst{\ref{aff16}}
\and A.~Veropalumbo\orcid{0000-0003-2387-1194}\inst{\ref{aff23},\ref{aff33},\ref{aff32}}
\and Y.~Wang\orcid{0000-0002-4749-2984}\inst{\ref{aff102}}
\and J.~Weller\orcid{0000-0002-8282-2010}\inst{\ref{aff31},\ref{aff15}}
\and G.~Zamorani\orcid{0000-0002-2318-301X}\inst{\ref{aff24}}
\and F.~M.~Zerbi\orcid{0000-0002-9996-973X}\inst{\ref{aff23}}
\and I.~A.~Zinchenko\orcid{0000-0002-2944-2449}\inst{\ref{aff116}}
\and E.~Zucca\orcid{0000-0002-5845-8132}\inst{\ref{aff24}}
\and M.~Sereno\orcid{0000-0003-0302-0325}\inst{\ref{aff24},\ref{aff29}}}
										   
%%%% please do not edit the affiliation list -- contact ECEB Bureau for changes
\institute{INAF - Osservatorio Astronomico d'Abruzzo, Via Maggini, 64100, Teramo, Italy\label{aff1}
\and
National Centre for Nuclear Research, ul. Pasteura 7, 02-093, Warsaw, Poland\label{aff2}
\and
Gran Sasso Science Institute (GSSI), Viale F. Crispi 7, L'Aquila (AQ), 67100, Italy\label{aff3}
\and
Universit\'e Paris Cit\'e, CNRS, Astroparticule et Cosmologie, 75013 Paris, France\label{aff4}
\and
CNRS-UCB International Research Laboratory, Centre Pierre Bin\'etruy, IRL2007, CPB-IN2P3, Berkeley, USA\label{aff5}
\and
Universit\'e de Strasbourg, CNRS, Observatoire astronomique de Strasbourg, UMR 7550, 67000 Strasbourg, France\label{aff6}
\and
Institute of Astronomy, University of Cambridge, Madingley Road, Cambridge CB3 0HA, UK\label{aff7}
\and
INAF-Osservatorio Astronomico di Roma, Via Frascati 33, 00078 Monteporzio Catone, Italy\label{aff8}
\and
INAF-Osservatorio Astrofisico di Arcetri, Largo E. Fermi 5, 50125, Firenze, Italy\label{aff9}
\and
Universit\'e Paris-Saclay, Universit\'e Paris Cit\'e, CEA, CNRS, AIM, 91191, Gif-sur-Yvette, France\label{aff10}
\and
NSF's NOIR, Lab 950 N. Cherry Avenue, Tucson, Arizona 85719, USA\label{aff11}
\and
Steward Observatory, University of Arizona, 933 N. Cherry Ave, Tucson, AZ 85750, USA\label{aff12}
\and
Utah Valley University, Orem, UT 84058, USA\label{aff13}
\and
INAF-Osservatorio Astronomico di Trieste, Via G. B. Tiepolo 11, 34143 Trieste, Italy\label{aff14}
\and
Max Planck Institute for Extraterrestrial Physics, Giessenbachstr. 1, 85748 Garching, Germany\label{aff15}
\and
Kapteyn Astronomical Institute, University of Groningen, PO Box 800, 9700 AV Groningen, The Netherlands\label{aff16}
\and
Universit\"at Innsbruck, Institut f\"ur Astro- und Teilchenphysik, Technikerstr. 25/8, 6020 Innsbruck, Austria\label{aff17}
\and
Space physics and astronomy research unit, University of Oulu, Pentti Kaiteran katu 1, FI-90014 Oulu, Finland\label{aff18}
\and
Isaac Newton Group of Telescopes, Apartado 321, 38700 Santa Cruz de La Palma, Spain\label{aff19}
\and
Max-Planck-Institut f\"ur Astronomie, K\"onigstuhl 17, 69117 Heidelberg, Germany\label{aff20}
\and
School of Physics and Astronomy, University of Nottingham, University Park, Nottingham NG7 2RD, UK\label{aff21}
\and
ESAC/ESA, Camino Bajo del Castillo, s/n., Urb. Villafranca del Castillo, 28692 Villanueva de la Ca\~nada, Madrid, Spain\label{aff22}
\and
INAF-Osservatorio Astronomico di Brera, Via Brera 28, 20122 Milano, Italy\label{aff23}
\and
INAF-Osservatorio di Astrofisica e Scienza dello Spazio di Bologna, Via Piero Gobetti 93/3, 40129 Bologna, Italy\label{aff24}
\and
IFPU, Institute for Fundamental Physics of the Universe, via Beirut 2, 34151 Trieste, Italy\label{aff25}
\and
INFN, Sezione di Trieste, Via Valerio 2, 34127 Trieste TS, Italy\label{aff26}
\and
SISSA, International School for Advanced Studies, Via Bonomea 265, 34136 Trieste TS, Italy\label{aff27}
\and
Dipartimento di Fisica e Astronomia, Universit\`a di Bologna, Via Gobetti 93/2, 40129 Bologna, Italy\label{aff28}
\and
INFN-Sezione di Bologna, Viale Berti Pichat 6/2, 40127 Bologna, Italy\label{aff29}
\and
INAF-Osservatorio Astronomico di Padova, Via dell'Osservatorio 5, 35122 Padova, Italy\label{aff30}
\and
Universit\"ats-Sternwarte M\"unchen, Fakult\"at f\"ur Physik, Ludwig-Maximilians-Universit\"at M\"unchen, Scheinerstr.~1, 81679 M\"unchen, Germany\label{aff31}
\and
Dipartimento di Fisica, Universit\`a di Genova, Via Dodecaneso 33, 16146, Genova, Italy\label{aff32}
\and
INFN-Sezione di Genova, Via Dodecaneso 33, 16146, Genova, Italy\label{aff33}
\and
Department of Physics "E. Pancini", University Federico II, Via Cinthia 6, 80126, Napoli, Italy\label{aff34}
\and
INAF-Osservatorio Astronomico di Capodimonte, Via Moiariello 16, 80131 Napoli, Italy\label{aff35}
\and
Dipartimento di Fisica, Universit\`a degli Studi di Torino, Via P. Giuria 1, 10125 Torino, Italy\label{aff36}
\and
INFN-Sezione di Torino, Via P. Giuria 1, 10125 Torino, Italy\label{aff37}
\and
INAF-Osservatorio Astrofisico di Torino, Via Osservatorio 20, 10025 Pino Torinese (TO), Italy\label{aff38}
\and
INAF-IASF Milano, Via Alfonso Corti 12, 20133 Milano, Italy\label{aff39}
\and
INFN-Sezione di Roma, Piazzale Aldo Moro, 2 - c/o Dipartimento di Fisica, Edificio G. Marconi, 00185 Roma, Italy\label{aff40}
\and
Centro de Investigaciones Energ\'eticas, Medioambientales y Tecnol\'ogicas (CIEMAT), Avenida Complutense 40, 28040 Madrid, Spain\label{aff41}
\and
Port d'Informaci\'{o} Cient\'{i}fica, Campus UAB, C. Albareda s/n, 08193 Bellaterra (Barcelona), Spain\label{aff42}
\and
Institute for Theoretical Particle Physics and Cosmology (TTK), RWTH Aachen University, 52056 Aachen, Germany\label{aff43}
\and
Deutsches Zentrum f\"ur Luft- und Raumfahrt e. V. (DLR), Linder H\"ohe, 51147 K\"oln, Germany\label{aff44}
\and
INFN section of Naples, Via Cinthia 6, 80126, Napoli, Italy\label{aff45}
\and
Institute for Astronomy, University of Hawaii, 2680 Woodlawn Drive, Honolulu, HI 96822, USA\label{aff46}
\and
Dipartimento di Fisica e Astronomia "Augusto Righi" - Alma Mater Studiorum Universit\`a di Bologna, Viale Berti Pichat 6/2, 40127 Bologna, Italy\label{aff47}
\and
Instituto de Astrof\'{\i}sica de Canarias, E-38205 La Laguna, Tenerife, Spain\label{aff48}
\and
Institute for Astronomy, University of Edinburgh, Royal Observatory, Blackford Hill, Edinburgh EH9 3HJ, UK\label{aff49}
\and
Jodrell Bank Centre for Astrophysics, Department of Physics and Astronomy, University of Manchester, Oxford Road, Manchester M13 9PL, UK\label{aff50}
\and
European Space Agency/ESRIN, Largo Galileo Galilei 1, 00044 Frascati, Roma, Italy\label{aff51}
\and
Universit\'e Claude Bernard Lyon 1, CNRS/IN2P3, IP2I Lyon, UMR 5822, Villeurbanne, F-69100, France\label{aff52}
\and
Institut de Ci\`{e}ncies del Cosmos (ICCUB), Universitat de Barcelona (IEEC-UB), Mart\'{i} i Franqu\`{e}s 1, 08028 Barcelona, Spain\label{aff53}
\and
Instituci\'o Catalana de Recerca i Estudis Avan\c{c}ats (ICREA), Passeig de Llu\'{\i}s Companys 23, 08010 Barcelona, Spain\label{aff54}
\and
Institut de Ciencies de l'Espai (IEEC-CSIC), Campus UAB, Carrer de Can Magrans, s/n Cerdanyola del Vall\'es, 08193 Barcelona, Spain\label{aff55}
\and
UCB Lyon 1, CNRS/IN2P3, IUF, IP2I Lyon, 4 rue Enrico Fermi, 69622 Villeurbanne, France\label{aff56}
\and
Mullard Space Science Laboratory, University College London, Holmbury St Mary, Dorking, Surrey RH5 6NT, UK\label{aff57}
\and
Department of Astronomy, University of Geneva, ch. d'Ecogia 16, 1290 Versoix, Switzerland\label{aff58}
\and
Universit\'e Paris-Saclay, CNRS, Institut d'astrophysique spatiale, 91405, Orsay, France\label{aff59}
\and
Aix-Marseille Universit\'e, CNRS/IN2P3, CPPM, Marseille, France\label{aff60}
\and
INAF-Istituto di Astrofisica e Planetologia Spaziali, via del Fosso del Cavaliere, 100, 00100 Roma, Italy\label{aff61}
\and
School of Physics, HH Wills Physics Laboratory, University of Bristol, Tyndall Avenue, Bristol, BS8 1TL, UK\label{aff62}
\and
University Observatory, LMU Faculty of Physics, Scheinerstr.~1, 81679 Munich, Germany\label{aff63}
\and
FRACTAL S.L.N.E., calle Tulip\'an 2, Portal 13 1A, 28231, Las Rozas de Madrid, Spain\label{aff64}
\and
Institute of Theoretical Astrophysics, University of Oslo, P.O. Box 1029 Blindern, 0315 Oslo, Norway\label{aff65}
\and
Leiden Observatory, Leiden University, Einsteinweg 55, 2333 CC Leiden, The Netherlands\label{aff66}
\and
Jet Propulsion Laboratory, California Institute of Technology, 4800 Oak Grove Drive, Pasadena, CA, 91109, USA\label{aff67}
\and
Department of Physics, Lancaster University, Lancaster, LA1 4YB, UK\label{aff68}
\and
Felix Hormuth Engineering, Goethestr. 17, 69181 Leimen, Germany\label{aff69}
\and
Technical University of Denmark, Elektrovej 327, 2800 Kgs. Lyngby, Denmark\label{aff70}
\and
Cosmic Dawn Center (DAWN), Denmark\label{aff71}
\and
NASA Goddard Space Flight Center, Greenbelt, MD 20771, USA\label{aff72}
\and
Department of Physics and Astronomy, University College London, Gower Street, London WC1E 6BT, UK\label{aff73}
\and
Universit\'e de Gen\`eve, D\'epartement de Physique Th\'eorique and Centre for Astroparticle Physics, 24 quai Ernest-Ansermet, CH-1211 Gen\`eve 4, Switzerland\label{aff74}
\and
Department of Physics, P.O. Box 64, University of Helsinki, 00014 Helsinki, Finland\label{aff75}
\and
Helsinki Institute of Physics, Gustaf H{\"a}llstr{\"o}min katu 2, University of Helsinki, 00014 Helsinki, Finland\label{aff76}
\and
Laboratoire d'etude de l'Univers et des phenomenes eXtremes, Observatoire de Paris, Universit\'e PSL, Sorbonne Universit\'e, CNRS, 92190 Meudon, France\label{aff77}
\and
SKAO, Jodrell Bank, Lower Withington, Macclesfield SK11 9FT, UK\label{aff78}
\and
Centre de Calcul de l'IN2P3/CNRS, 21 avenue Pierre de Coubertin 69627 Villeurbanne Cedex, France\label{aff79}
\and
Universit\"at Bonn, Argelander-Institut f\"ur Astronomie, Auf dem H\"ugel 71, 53121 Bonn, Germany\label{aff80}
\and
Aix-Marseille Universit\'e, CNRS, CNES, LAM, Marseille, France\label{aff81}
\and
Dipartimento di Fisica e Astronomia "Augusto Righi" - Alma Mater Studiorum Universit\`a di Bologna, via Piero Gobetti 93/2, 40129 Bologna, Italy\label{aff82}
\and
Department of Physics, Institute for Computational Cosmology, Durham University, South Road, Durham, DH1 3LE, UK\label{aff83}
\and
University of Applied Sciences and Arts of Northwestern Switzerland, School of Engineering, 5210 Windisch, Switzerland\label{aff84}
\and
Institute of Physics, Laboratory of Astrophysics, Ecole Polytechnique F\'ed\'erale de Lausanne (EPFL), Observatoire de Sauverny, 1290 Versoix, Switzerland\label{aff85}
\and
Telespazio UK S.L. for European Space Agency (ESA), Camino bajo del Castillo, s/n, Urbanizacion Villafranca del Castillo, Villanueva de la Ca\~nada, 28692 Madrid, Spain\label{aff86}
\and
Institut de F\'{i}sica d'Altes Energies (IFAE), The Barcelona Institute of Science and Technology, Campus UAB, 08193 Bellaterra (Barcelona), Spain\label{aff87}
\and
School of Mathematics and Physics, University of Surrey, Guildford, Surrey, GU2 7XH, UK\label{aff88}
\and
European Space Agency/ESTEC, Keplerlaan 1, 2201 AZ Noordwijk, The Netherlands\label{aff89}
\and
School of Mathematics, Statistics and Physics, Newcastle University, Herschel Building, Newcastle-upon-Tyne, NE1 7RU, UK\label{aff90}
\and
DARK, Niels Bohr Institute, University of Copenhagen, Jagtvej 155, 2200 Copenhagen, Denmark\label{aff91}
\and
Waterloo Centre for Astrophysics, University of Waterloo, Waterloo, Ontario N2L 3G1, Canada\label{aff92}
\and
Department of Physics and Astronomy, University of Waterloo, Waterloo, Ontario N2L 3G1, Canada\label{aff93}
\and
Perimeter Institute for Theoretical Physics, Waterloo, Ontario N2L 2Y5, Canada\label{aff94}
\and
Space Science Data Center, Italian Space Agency, via del Politecnico snc, 00133 Roma, Italy\label{aff95}
\and
Centre National d'Etudes Spatiales -- Centre spatial de Toulouse, 18 avenue Edouard Belin, 31401 Toulouse Cedex 9, France\label{aff96}
\and
Institute of Space Science, Str. Atomistilor, nr. 409 M\u{a}gurele, Ilfov, 077125, Romania\label{aff97}
\and
Consejo Superior de Investigaciones Cientificas, Calle Serrano 117, 28006 Madrid, Spain\label{aff98}
\and
Universidad de La Laguna, Dpto. Astrof\'\i sica, E-38206 La Laguna, Tenerife, Spain\label{aff99}
\and
Dipartimento di Fisica e Astronomia "G. Galilei", Universit\`a di Padova, Via Marzolo 8, 35131 Padova, Italy\label{aff100}
\and
INFN-Padova, Via Marzolo 8, 35131 Padova, Italy\label{aff101}
\and
Caltech/IPAC, 1200 E. California Blvd., Pasadena, CA 91125, USA\label{aff102}
\and
Instituto de F\'isica Te\'orica UAM-CSIC, Campus de Cantoblanco, 28049 Madrid, Spain\label{aff103}
\and
Institut de Recherche en Astrophysique et Plan\'etologie (IRAP), Universit\'e de Toulouse, CNRS, UPS, CNES, 14 Av. Edouard Belin, 31400 Toulouse, France\label{aff104}
\and
Universit\'e St Joseph; Faculty of Sciences, Beirut, Lebanon\label{aff105}
\and
Departamento de F\'isica, FCFM, Universidad de Chile, Blanco Encalada 2008, Santiago, Chile\label{aff106}
\and
Department of Physics and Helsinki Institute of Physics, Gustaf H\"allstr\"omin katu 2, University of Helsinki, 00014 Helsinki, Finland\label{aff107}
\and
Infrared Processing and Analysis Center, California Institute of Technology, Pasadena, CA 91125, USA\label{aff108}
\and
Departamento de F\'isica, Faculdade de Ci\^encias, Universidade de Lisboa, Edif\'icio C8, Campo Grande, PT1749-016 Lisboa, Portugal\label{aff109}
\and
Instituto de Astrof\'isica e Ci\^encias do Espa\c{c}o, Faculdade de Ci\^encias, Universidade de Lisboa, Tapada da Ajuda, 1349-018 Lisboa, Portugal\label{aff110}
\and
Cosmic Dawn Center (DAWN)\label{aff111}
\and
Niels Bohr Institute, University of Copenhagen, Jagtvej 128, 2200 Copenhagen, Denmark\label{aff112}
\and
Universidad Polit\'ecnica de Cartagena, Departamento de Electr\'onica y Tecnolog\'ia de Computadoras,  Plaza del Hospital 1, 30202 Cartagena, Spain\label{aff113}
\and
Institute of Space Sciences (ICE, CSIC), Campus UAB, Carrer de Can Magrans, s/n, 08193 Barcelona, Spain\label{aff114}
\and
Institut d'Estudis Espacials de Catalunya (IEEC),  Edifici RDIT, Campus UPC, 08860 Castelldefels, Barcelona, Spain\label{aff115}
\and
Astronomisches Rechen-Institut, Zentrum f\"ur Astronomie der Universit\"at Heidelberg, M\"onchhofstr. 12-14, 69120 Heidelberg, Germany\label{aff116}}

 \abstract{

Surface brightness fluctuations (SBF), the pixel-to-pixel variations in flux that arise from the statistical distribution of stars in galaxy images, provide a powerful tool to measure redshift-independent distances from photometric data alone. This method is particularly important in the era of large imaging surveys, such as those carried out during the \Euclid mission. Here we present the first application of SBF to \Euclid data, using the {\texttt{FAST-SBF}} code to measure stellar fluctuation amplitudes in the \IE band for a sample of galaxies in the Early Release Observations (ERO) of the Fornax galaxy cluster. Although the \Euclid data reduction pipeline is not optimized for SBF measurements, extensive testing suggests that we are able to extract robust results. We calibrate the absolute fluctuation magnitude $\overline{M}_\mathrm{IE}$ as a function of the $(\IE{-}\HE)$ colour for 15 galaxies in Fornax, then test this relation on galaxies in two other ERO fields: the Perseus cluster and dwarf satellite candidates around NGC\,6744. Overall, we find reasonable agreement with distances in the literature and good agreement in cases where {\texttt{FAST-SBF} indicates the measurements are robust.} Finally, we compare our results against Stellar Population Tools (SPoT) simple stellar population models, and discuss the possibility of using SBF with \Euclid to probe the underlying stellar populations of the galaxies.  }
  

   \titlerunning{\Euclid\/: Calibrating SBF distances in Fornax}
   \authorrunning{R. Habas et al.}
   
   \maketitle

\section{\label{sc:intro}Introduction}

By the end of its six years of planned operations, the \Euclid mission will have observed approximately \num{14000}\,deg$^2$ of the sky as part of the Euclid Wide Survey (EWS) and several deep and medium-deep fields, imaging billions of galaxies at visible and near-infrared (NIR) wavelengths \citep{Scaramella-EP1}. The spatial and wavelength coverage, resolution, depth, and low background will make the survey data ideal for any number of science projects, including efforts to map the 3D structure of the nearby Universe using surface brightness fluctuations (SBF). Before such an ambitious project can commence, however, it is necessary to standardise the method, essentially calibrating the distance zero point, for the \Euclid passbands.

Among the various local distance indicators, SBF is particularly versatile. The SBF signal arises from the discrete nature of star counts; pixel-to-pixel variations in flux are generated in the light profile of a galaxy due to the purely statistical variation in the number of stars that fall on a given detector pixel \citep{Tonry1997,Tonry1990,Tonry1988}. The amplitude of the power spectrum of these fluctuations scales with $\propto d^{-2}$, where d is the galaxy distance, providing a constraint on the distance of a galaxy from imaging alone, once an appropriate calibration of the absolute SBF amplitude has been determined. The current precision of the method is approximately 5\% for bright, massive elliptical galaxies \citep{Cantiello2024b}, but is more typically 10\%--20\% for dwarf galaxies \citep[e.g.,][]{Jensen2015}. Nevertheless, SBF is still a powerful tool for dwarfs, for which SBF magnitudes may represent the only available distance indicator  \citep{Cantiello2024b,Moresco2022}. The technique has been successfully applied to galaxies at distances exceeding 100\,Mpc using \HST imaging \citep{Jensen2025,Blakeslee2021,Biscardi2008,Jensen2003,Jensen2001}.

One caveat of the method is that SBF is traditionally applied to early-type galaxies (ETGs). The fluctuations of interest, arising from stellar counts, can be affected by dust and star-forming regions, or confused with GC, foreground stars, background galaxies, and nearby extended objects. In addition, morphological irregularities and mixed stellar populations influence the pixel-by-pixel Poissonian fluctuations.  Contamination from these sources can be minimized by working with ETGs, which typically have smooth morphologies, host older stellar populations, and have low dust content. However, it can also be applied to late-type galaxies (LTGs) if a relatively smooth region, free of dust and ongoing star formation, can be identified \citep[e.g.,][]{Cantiello2024,Tonry2001,Tonry1997}. The technique has also been successfully applied to dwarf galaxies \citep[e.g.,][]{Kim2021,Carlsten2019,Blakeslee2009,Mei2007,Jerjen2003}, including ultra-diffuse galaxies such as NGC\,1052-DF2 \citep{vanDokkum2018,Blakeslee2018}.

It is well established in the literature that, at a fixed distance, the SBF amplitude has a colour dependence due to the properties of the underlying stellar population in the region where the signal is measured \citep[e.g.,][]{Vazdekis2020,Blakeslee2009b,Cantiello2007}. At a given age, metal-rich populations exhibit lower amplitude (fainter) fluctuations than metal-poor populations. Similarly, at a fixed metallicity, older stellar populations have lower luminosity fluctuations than younger populations. This colour dependence means that distances cannot be extracted from the fluctuation amplitude alone, as the stellar population dependence of the apparent SBF magnitude $\overline{m}$  needs to be calibrated, and a zero point derived. This classical procedure is usually referred to as the standardization of the distance indicator. 

Given the spatial coverage, photometric uniformity, and image quality of \Euclid, we expect the telescope will provide optimal imaging to measure SBF for thousands of galaxies, and possibly more. This paper will test the measurability of SBF with \Euclid data, and generate a first template calibration for SBF in the \Euclid \IE band using ERO imaging of the Fornax galaxy cluster. 

Fornax is a nearby, well studied cluster that has been observed across the wavelength spectrum \citep[e.g.,][]{Serra2023,Saifollahi21,Sarzi2018,Pota2018,Iodice2016}, and into the dwarf regime \citep[e.g.,][]{Zabel2024,Kleiner2023,Su2021,Venhola2019,Eigenthaler2018}. The cluster appears to be in an intermediate dynamical state, with some evidence for recent infalling galaxies but a relatively relaxed cluster core \citep{Zhu2022,Iodice2019,Sheardown2018}. The cluster is also fairly compact, with a line-of-sight depth of $2.0^{+0.4}_{-0.6}$\,Mpc \citep{Blakeslee2009}, and, aside from the Fornax A subgroup, is isolated in velocity space \citep{Maddox2019}; taken together, this implies that this is an ideal cluster to test the intrinsic scatter in the SBF method. The distances to individual galaxies in the cluster have been estimated using various methods, including SBF, the planetary nebula luminosity function, and the tip of the red giant branch \citep[TRGB; e.g.,][]{Anand2024,Spriggs2021,Blakeslee2009}. For the cluster itself, we adopt the TRGB distance $d = (19.3 \pm 0.7)$\,Mpc from \citet{Anand2024}, based on recent \textit{James Webb} Space Telescope (JWST) observations.

TRGB measurements are ideal for standardizing the zeropoint of the SBF colour-magnitude relation. The TRGB is an accurate and precise standard candle \citep[e.g.,][]{Beaton2018,Serenelli2017}.  It also uses the same population (pop II) of stars that SBF probes, which should reduce any systematic uncertainties between the estimators \citep[e.g.,][]{Jensen2025,Anand2025,Anand2024}. Thus we will give preference to TRGB distances, where possible, throughout this work.

The paper is structured as follows: in Sect.\,\ref{sc:data}, we give an overview of the \Euclid imaging and the data reduction pipeline, and describe how our targets where selected; in Sect.\,\ref{sc:code}, we introduce the {\texttt{FAST-SBF}} code used for the SBF measurements; our results are presented in Sect.\,\ref{sc:results}, where we discuss the calibration of the \Euclid passbands and consistency tests to check the robustness of the results; finally, in Sect.\,\ref{sc:conclusions}, we summarize our results and provide a global outlook on the SBF method for \Euclid.

\section{\label{sc:data}Data}

The \Euclid ERO images were taken during the performance verification (PV) phase, shortly after the launch of the telescope in 2023. This programme included 17 fields that were selected to highlight the capabilities of \Euclid and its potential for non-cosmological science, ranging in distance from the Orion star forming region ($d\sim500$\,pc; \citealt{EROOrion}) to the Abell\,2390 galaxy cluster at $z=0.228$ \citep{EROLensData}. While the focus of this paper is verifying the measurability of SBF in \Euclid \IE images and standardizing the method using the Fornax ERO data \citep{EROFornaxGCs}, we will also use ERO imaging of the Perseus galaxy cluster \citep{EROPerseusOverview,EROPerseusDGs} and NGC\,6744 \citep{ERONearbyGals} to test our subsequent calibration. 

To maximize the scientific return of these fields, the observing strategy of the ERO targets was allowed to differ somewhat from the standard procedure that is adopted for the full EWS, and between ERO targets. In general, EWS observations are grouped into reference observing sequences (ROS) where each ROS includes four dithers of a single field (0.53\,deg$^2$) and the associated calibration data for all four filters (\IE, \YE, \JE, and \HE) on the VIS \citep{EuclidSkyVIS} and NISP \citep{EuclidSkyNISP} instruments. Some ERO fields, however, were imaged with more than one ROS and they can cover larger footprints than a standard EWS tile. Full details are provided by \citet{EROData}, but we highlight key details below.

\begin{figure*}[htbp!]
  \begin{center}
    \includegraphics[width=1.95\columnwidth]{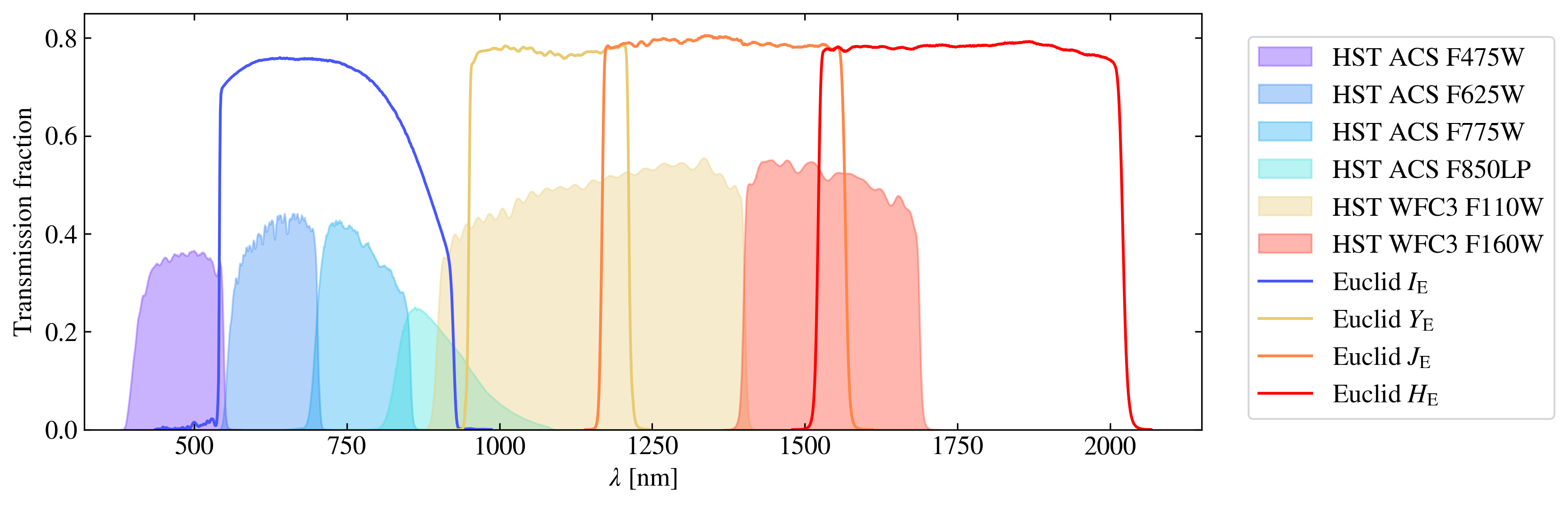}
   \end{center}
\caption{Transmission filters for \Euclid compared to selected HST filters, which have been used for prior SBF measurements. \citet{Blakeslee2009} used HST/ACS $g_\mathrm{F475W}$ and $z_\mathrm{F850LP}$ observations to measure SBF distances for eight of the galaxies in our sample. Other studies have used the WFC3 F110W and F160W filters to measure SBF in the NIR \citep{Jensen21,Jensen2015}. All transmission curves are from the Spanish Virtual observatory Filter Profile Service \citep{Rodrigo2020}; this NISP filter curves can also be found in the paper by \citet{Schirmer-EP18}. }
\label{fig:transmission}
\end{figure*}

The ERO images were also reduced using an alternative to the science ground segment (SGS) data reduction pipeline. 
In particular, we use the `low surface brightness stack' images, which preserve the extended emission around both high- and low surface brightness sources \citep{EROData}. Subsequent tests indicate that the photometric accuracy of the ERO reductions is within a few percent in each band. Unless otherwise noted, all magnitudes are given in the AB system.

One difference between the ERO and SGS pipelines that is key for SBF measurements is the choice of interpolation kernel that is used for stacking and rotating individual frames. The Lanczos family of kernels, particularly Lanczos3, is preferred for SBF analyses because it minimizes the correlation between pixels in the spatial power spectrum, and this technique assumes that the noise in each pixel is independent of all other pixels \citep[e.g.,][]{Mei2005IV}. The ERO pipeline applies a Lanczos3 kernel to the \IE images and a bilinear kernel to the three NIR bands.

In principle, the NISP passbands should be favoured for SBF measurements. In these filters, the SBF signal is brighter, the contrast with GC is greater, and the effects of dust are usually negligible (see \citealt{Jensen2015}, and references therein). Furthermore, the \IE band is so wide (see Fig.\,\ref{fig:transmission}), that the scatter due to stellar population effects could be larger than was previously measured in any optical passbands \citep[e.g.,][]{Cantiello2003,Blakeslee2001}. However, as noted above, the interpolation kernel used to stack the \IE images better preserves the noise characteristics of the spatial power spectrum. VIS also has a better pixel sampling (\ang{;;0.1} versus \ang{;;0.3}), which allows us to resolve fluctuations on smaller scales and provides more pixels for better statistics in the measurements. Thus, for this work we measure the SBF magnitudes in \IE ($\overline{m}_\mathrm{IE}$) and trace the underlying stellar populations using $(\IE{-}\HE)$ colours.

\subsection{\label{sc:ERO}Observations of the Fornax cluster and sample selection}
A portion of the Fornax cluster was imaged in August and September 2023 as part of the ERO observations. The 0.6\,deg$^2$ region, centred at RA\,$=\ang{54.017;;}$ and Dec\,$=-\ang{35.267;;}$, is offset from the cluster centre, but it does encompass several massive galaxies within the cluster core. A larger portion of the cluster will be included in Data Release 1 (DR1). 

Observations of the Fornax cluster suffered from a number of issues: early failures of the fine guidance sensor, initial issues with stray light, and a dither pattern that was mistakenly rotated by \ang{90;;}, which resulted in regions of zero coverage in the final stacked image \citep{EROData}. Nevertheless, two \IE exposures and four exposures in \YE, \JE, and \HE had sufficient quality for scientific use. The two \IE exposures are rotated \ang{1;;} and \ang{10;;} from north, respectively. In \IE and \HE, the two bands we will use in this work, the $5\sigma$ point source limits are $\IE=26.89$ and $\HE=24.81$, while the surface brightness depths are $\mu_{\IE}=29.66$\,mag\,arcsec$^{-2}$ and $\mu_{\HE}=28.56$\,mag\,arcsec$^{-2}$ \citep{EROData}.  

According to \citet{Borlaff-EP16}, the local component of the scattered light (e.g., on pixel scales) is uncertain. However, since it behaves as a white-noise component, any remaining flux is expected to affect only the signal to noise ratio (S/N) of the SBF signal, without altering the measurements themselves.

Our target sample is composed of nine well-known Fornax cluster members that have prior SBF distance estimates \citep{Blakeslee2009,Liu2002,Tonry2001}, supplemented by a sample of dwarf candidates compiled by a team within the Euclid Consortium. The methodology to identify the dwarfs is nearly identical to that presented by \citet{EROPerseusDGs}, where the ERO images were visually inspected by a subset of the co-authors and dwarf candidates were flagged using the online \texttt{JAFAR} \citep{Sola2025,Sola2022} tool. From this catalogue, we visually re-examined the dwarf candidates, keeping only those suitable for SBF measurements, i.e., dwarfs with smooth profiles that have coverage in both the \IE and \HE images, are large enough to have reasonable number of pixels for good statistics,\footnote{As a general rule of thumb, the annuli used for SBF measurements should have at least 1000 unmasked pixels for good statistics.} and are bright enough to easily distinguish them from the local background. The exact combination of parameters that makes a galaxy a good candidate for SBF distances cannot be easily quantified, so these are not strict parameter cuts. Based on this selection, we have a sample of 22 additional dwarf candidates, for a total of 31 galaxies. Twenty of the additional dwarf candidates are already known in the literature \citep{Venhola2022,Venhola2018,Eigenthaler2018,Paturel2003}, while two are new dwarf candidate identifications.

\subsection{\label{sc:ERO-Perseus}Observations of the Perseus cluster and sample selection}

The Perseus galaxy cluster was imaged with four ROS in order to mitigate regions of heavy extinction found in this area of the sky.  The final images should be 0.75\,mag\,arcsec$^{-2}$ deeper than the EWS, in theory, although the contamination from Galactic cirrus in the region mitigates most of this gain in practice. The common field of view (FoV) is centred on RA\,$=\ang{49.638;;}$ and Dec\,$=+\ang{41.651;;}$ and covers 0.7\,deg$^2$ of the cluster core. The orientation of the image is \ang{30;;} rotated clockwise from north. The observations only suffered from the rotated dither, again resulting in regions of reduced coverage. In \IE and \HE, the $5\sigma$ point source limits are $\IE=28.03$ and $\HE=25.32$, while the surface brightness depths are $\mu_{\IE}=30.57$\,mag\,arcsec$^{-2}$ and $\mu_{\HE}=28.94$\,mag\,arcsec$^{-2}$ \citep{EROData,EROPerseusDGs}.

Measuring SBF distances for a  number of galaxies in the Perseus cluster \citep{EROPerseusOverview,EROPerseusDGs} will be the subject of a future work. Here, we select a subsample of 19 galaxies to test (a) our ability to measure SBF at higher distances in \Euclid data and (b) the calibration that we derive for the Fornax sample. The galaxies were chosen based on a combination of parameter sampling and visual inspection, such that they are: relatively isolated, have smooth morphologies, are large enough to have good statistics, and span the range of measured colours in the cluster.

We adopt an aggregate distance of $d_{\mathrm{Perseus}} = (72 \pm 3)$\,Mpc to the cluster, based on various distance estimates for individual member galaxies reported in the literature (e.g., Type Ia supernovae, fundamental plane, Tully–Fisher, and SBF; \citealt{Tully2009}). This value will only be used to test our calibration line, and thus an averaged distance, which should reduce the systematic uncertainties between the measurements, is sufficiently accurate. The compactness of the cluster is not well constrained, but it is a massive cluster ($M_{200} = 1.2 \times 10^{15}\,M_{\odot}$) with a correspondingly large virial radius ($r_{200} = 2.2$\,Mpc) and velocity dispersion ($\sigma_{\mathrm{Perseus}} > 1000$\,km\,s$^{-1}$; \citealt{Aguerri2020}).

\subsection{\label{sc:ERONGC6744}Observations of NGC 6744 and sample selection}

NGC\,6744, a Milky Way analogue, is the most distant of the six targets in \citet{ERONearbyGals}, designed to illustrate the capabilities of \Euclid for science in the Local Volume. This field did not face any of the observing difficulties encountered by the other two targets. The galaxy was imaged with one standard ROS centred on the galaxy (RA$=\ang{287.430;;}$; Dec$=-\ang{63.842;;}$). The image covers 0.6\,deg$^2$ of sky and is rotated \ang{15;;} from north. In the two bands used for this work, the $5\sigma$ point source limits are $\IE=27.09$ and $\HE=24.58$, while the surface brightness depths are $\mu_{\IE}=29.81$\,mag\,arcsec$^{-2}$ and $\mu_{\HE}=28.21$\,mag\,arcsec$^{-2}$ \citep{EROData}.

As a nearly face-on SAB(r)bc galaxy, NGC\,6744 is not an ideal target for an SBF distance estimate; the galaxy is actively forming stars \citep{Yew2018} and there are numerous dust lanes throughout that would interfere with the SBF signal. Instead, we target the dwarf satellite candidates around the galaxy. \citet{Karachentsev2025} compiled a list of 25 tentative group members associated with NGC\,6744, of which six galaxies -- all dwarfs -- fall within the ERO footprint. \citet{Carlsten2022} have already used SBF to confirm that four of the six are dwarf satellites of NGC\,6744, classify one as a background dwarf, and were unable to determine a distance for the sixth. We use all six dwarfs to test our procedure and calibration on nearby, semi-resolved galaxies.

Two groups have recently measured the distance to NGC\,6744 using TRGB stars. \citet{Sabbi2018} reported a distance of 
$(8.8 \pm 0.8)$\,Mpc in an off-centred (northern) region of the galaxy. More recently, \citet{Anand2021} measured a distance of $(9.39 \pm 0.43)$\,Mpc. 

\begin{figure*}[htbp!]
  \begin{center}
    \includegraphics[width=1.98\columnwidth]{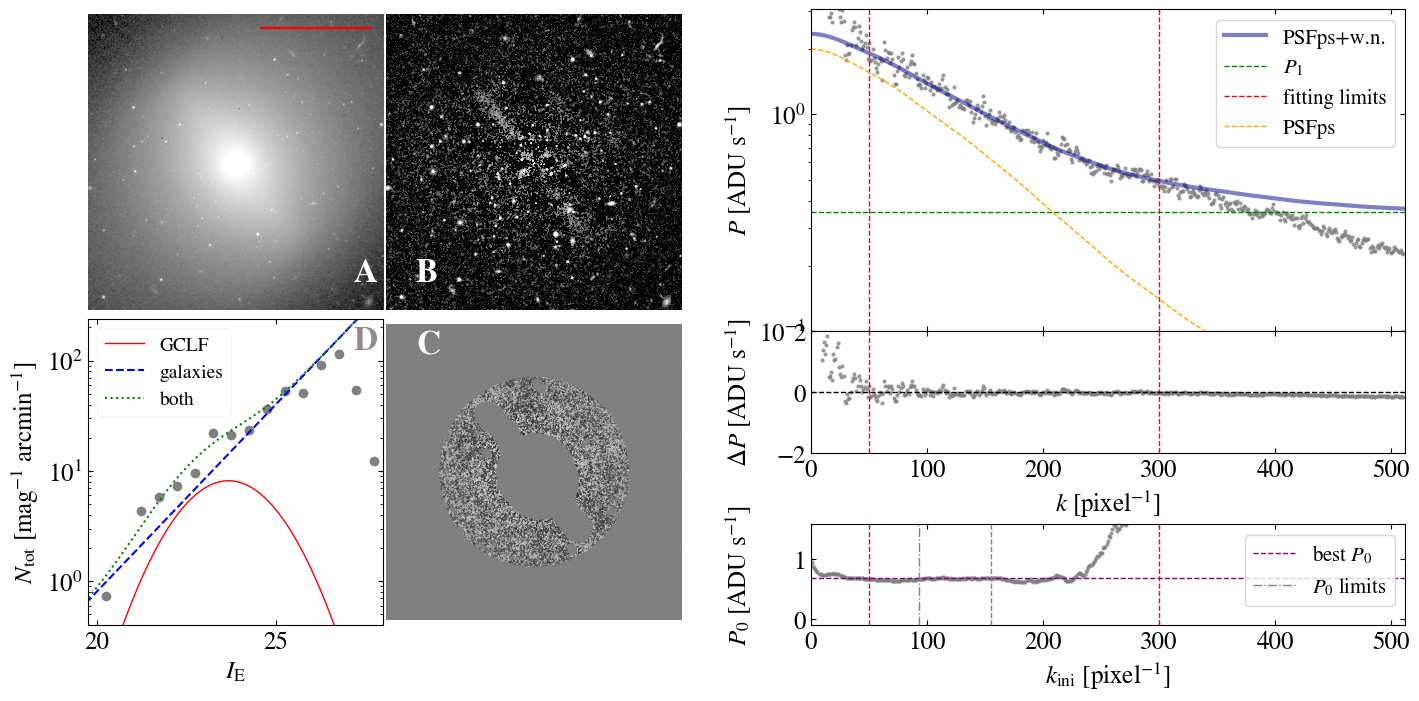}
   \end{center}
\caption{Key steps in the SBF process are shown for FCC\,190, a massive cluster member in the Fornax ERO image. \emph{Left panel, clockwise from upper left}: (a) the original \IE image of the galaxy; (b) the residual image after the background, galaxy model, and large-scale structure have been subtracted; (c) the annulus used for the measurements, with remaining spurious (non-SBF) sources masked; and (d) the fit to the combined luminosity function of GC and background galaxies. Note that the cutout size, which is fixed for the three images, was chosen for this display; {\texttt{FAST-SBF}} is run on a much larger cutout in order to accurately measure the background and model the outermost isophotes of the galaxy. The red bar in the upper right corner of figure (a) represents \ang{;;30}. North is up and east is to the left. \emph{Right panel, from top to bottom}: (i) the azimuthally averaged radial power spectrum profile, where the data is plotted in grey, the best-fit line is shown in blue and includes contributions from the power spectrum of the PSF (labelled PSFps; yellow dashed line) and a white noise (w.n.) component ($P_\mathrm{1}$; green), while the bounds of the fitting region are marked with red dashed lines; (ii) the difference between the model and the best-fit line as a function of the wavenumber $k$; and (iii) the amplitude $P_\mathrm{0}$ that would be extracted from the fit, if we iterate over the fitting region, increasing $k_\mathrm{ini}$ by one data point each time. The best $P_\mathrm{0}$ value is determined from the most stable value within a specified fraction of the total window fitting region, marked with black dashed lines. Similar plots for other massive galaxies in the cluster are shown in Appendix\,\ref{appendix:fits}. }
\label{fig:process_example1}
\end{figure*}

\section{\label{sc:code}SBF measurements}
To isolate the SBF signal and measure its amplitude, we follow the philosophy described in several prior works \citep[e.g.,][]{Cantiello2024,Moresco2022,Blakeslee2021,Jensen21,Mei2005V}. 
In brief, for a single galaxy, this procedure involves: (a) estimating and subtracting the local sky background; (b) modelling the galaxy and subtracting it from the image; (c) removing any remaining large-scale residuals; (d) identifying compact and extended sources to be masked and estimating, from their luminosity function, the fluctuations arising from undetected GC and background galaxies, respectively; (e) measuring a local point spread function (PSF); (f) converting the signal from the residual masked image into Fourier space, then measuring its amplitude from the azimuthally averaged radial power spectrum by matching the residual-frame power spectrum to that of the PSF; (g) measuring the colour in the same masked annulus from which the SBF amplitude is measured; and (h) calculating the SBF magnitude $\overline{m}_\mathrm{IE}$ and, based on its position in $\overline{m}$ versus colour space, estimating the absolute SBF magnitude $\overline{M}_\mathrm{IE}$ to derive the distance modulus of the galaxy.

We use a {\texttt{python}} implementation of this procedure, the {\texttt{Flexible Automated Self-contained Tool for SBF}} (\texttt{FAST-SBF}) package,\footnote{{\texttt{FAST-SBF}} will be released to the wider community after planned improvements have been implemented. Anyone interested in early access to the code is welcome to contact either Michele Cantiello (michele.cantiello@inaf.it) or Rebecca Habas (rebecca.habas@inaf.it).} first described by \citet{Riccio2025}. The code is still under development, but the key modules are in place and the output is consistent with SBF measurements in the literature. It is designed to be self-consistent, largely automated, and will work with optical or NIR images from any telescope. Manual intervention has been limited to setting galaxy-specific parameters in a configuration file, and optional checkpoints where users can visually inspect the images and manually mask (via an interactive graphical user interface) any sources that were not fully removed by the automated routines but could still affect the galaxy modelling or SBF measurements. The only script that is not native to {\texttt{python}} is {\texttt{Source Extractor}} \citep{Bertin1996}; the full version is called from within {\texttt{FAST-SBF}}, so a local installation is required. 

In order to run, {\texttt{FAST-SBF}} requires the aforementioned configuration file as well as science, mask, and weight images in two bands. The images, or image cutouts, should be large enough to estimate the local sky background, model the outermost galaxy isophotes, and contain enough stars to construct a local PSF. 

Although PSFs are estimated at various points in the official \Euclid pipeline, \texttt{FAST-SBF} has been designed to create an independent estimate of the PSF, to make it versatile enough to work with other telescopes and to account for any local peculiarities in the imaging. A catalogue of candidate PSF reference stars is built from {\texttt{Source Extractor}} parameters, based on user-specified magnitude cuts, isolation criteria, and compactness. 

A more detailed description of {\texttt{FAST-SBF}} is presented by \citet{Riccio2025}, including a flowchart of the pipeline (see their figure 2). Below, we highlight some of the key modifications needed to apply the code to \Euclid images. 

\subsection{Background subtraction}
The final \Euclid images, described above, have had flat-field corrections applied and a global background subtracted. Nevertheless, it is important to remove any remaining residual or local sky background, as this emission can impact the SBF magnitudes, colours, and associated errors. 

The background in Fornax is expected to be relatively flat, and indeed we measure less than a 2\% variation in the sky background for all of our targets in the \IE band, although this rises to 7\% in \HE. We test the effect of this uncertainty on our final values, assuming a $\pm5\%$ uncertainty in the background of both filters. For the bright galaxies, there is little impact on the measured parameters, with $\Delta \overline{m}_\mathrm{IE} \lesssim 0.01$, and $\Delta(\IE{-}\HE) \lesssim 0.015$, which is well within the anticipated uncertainties. In general, this is consistent with existing results, as the sky uncertainty is expected to contribute $\leq 0.02$ to the total error on $\overline{m}$, corresponding to $<1\%$ of the final distance \citep{Cantiello2024b}. Dwarf galaxies show similarly small $\Delta \overline{m}_\mathrm{IE}$ values for a $\pm5\%$ uncertainty in the background, but their colours are more sensitive to errors and can change by $\Delta(\IE{-}\HE) \lesssim 0.25$.

The Fornax cluster also has intracluster light (ICL) extending from NGC\,1399 in the cluster centre (not in the Fornax ERO footprint), to NGC\,1381, NGC\,1387, and NGC\,1379 \citep{Iodice2017,Iodice2016}. In the \Euclid ERO images, this covers roughly a quarter of the image, and is primarily restricted to one quadrant \citep{EROFornaxGCs}. The surface brightness of the ICL is on the order of $\mu_r \simeq (28$\,--\,$29$)\,mag\,arcsec$^{-2}$, as measured in the Very large telescope Survey Telescope (VST) $r$-band \citep{Iodice2017}. It is not expected to have a significant impact on the SBF or colour estimates, but we will explicitly test this assumption in Sect. \ref{sc:SBFsky}.

In the current version of the code, {\texttt{FAST-SBF}} estimates and subtracts a constant sky background value. This does not accurately represent the background for dwarf galaxies near contaminating sources (e.g., bright companions, stars, and image artifacts). Indeed, this, combined with the higher sensitivity to sky-background uncertainties mentioned above, represents a problem for several of the Fornax dwarfs in this sample, as discussed in Sects.\,\ref{sc:SBFsky}--\ref{sc:calibration}.

\subsection{\label{sec:spurious}Spurious fluctuation sources}
From the initial conception of the SBF technique, it has been known that GC and background galaxies, which we collectively refer to as `spurious sources,' contaminate the SBF signal \citep[e.g.,][]{Tonry1990,Tonry1989,Tonry1988}. This problem is particularly pronounced in massive galaxies, which usually host large numbers of GC and subtend more background galaxies due to their larger apparent sizes. We can remove the spurious fluctuations by first masking all GC and background galaxies above a given S/N. These masked sources are then fit with a combined luminosity function, composed of a Gaussian distribution for the GC \citep{Lomeli-Nunez2022,Rejkuba2012,Whitmore1995,Reed1994} and a power law for the background galaxies, which allows us to also estimate the fluctuation term from unmasked (faint) spurious sources. The contribution of these sources to the power spectrum amplitude will be denoted by $P_\mathrm{r}$ \citep[see][and Sect. \ref{sec:sbf_mags}]{Cantiello2005}.

To fit the GC component of the composite luminosity function, we use the characterisation of the GCLF in the Fornax ERO imaging as a prior. For the massive galaxies in the cluster, \citet{EROFornaxGCs} found that the turnover magnitude (TOM) occurs at $\IE = 22.90 \pm 0.09$, or an absolute magnitude ${\rm{TOM}} = -8.59 \pm 0.09$. The limiting magnitude of the image is several orders of magnitude fainter, which means that that GCLF should be well sampled past the peak, and we can estimate the contribution of undetected GC with reasonable confidence. We further adopt $\sigma_{\rm{GCLF}} = 1.2$ for the width of the distribution \citep{Villegas2010}. 

 For the dwarf galaxies, we identify and mask any GC above the detection threshold, but do not attempt to model the GCLF to correct for fainter sources; although the specific frequency is high \citep[e.g.,][]{Marleau2021,Georgiev2010}, the number of GC is so low that fitting a Gaussian luminosity function to those few data points becomes highly unreliable. The contribution of undetected GC is almost certainly less than the uncertainty in the SBF signal in dwarfs, and a visual inspection of the images verifies that we have no dwarfs with a-typically large GC populations -- such as MATLAS-2019 \citep{Marleau2024,Mueller2021,Mueller2020,Forbes2019}, NGC\,1052-DF2 \citep{Kahrion2025,vanDokkum2018b}, NGC\,1052-DF4 \citep{Shen2021}, or FCC\,224 \citep{Buzzo2025,Tang2025} -- in the sample.

The correction to account for background galaxy contamination is applied to all galaxies in the sample. For the background galaxy luminosity function, we assumed a power-law exponent $\gamma=0.34$ \citep{Cantiello2005}. Our results are not particularly sensitive to the exact value of this parameter because galaxies beyond the source detection limit are so faint that they contribute very little to the fluctuation signal.

It is particularly important for SBF measurements that contaminating sources are correctly distinguished from the mottling in the surface brightness of the galaxy that is generated by the stellar variance that we aim to measure. The SBF S/N typically increases toward the galaxy centre, to the point that the fluctuations can appear as real astrophysical objects. To avoid confusing physical sources with stellar fluctuations arising from Poisson statistics, source identification is performed using a modified weight map in which the galaxy model is multiplied by a scaling factor $\kappa$, and then added to the image weight map \citep[e.g.,][]{Jordan2004}. The value of $\kappa$ is band and observation dependent, and is currently determined empirically; in effect, the parameter raises the S/N threshold for source detection so that the fluctuations themselves are not masked as contaminants. After several iterations of the code, in which we balanced the requirements to mask all obvious compact and extended sources against leaving the brightness variance untouched in the region of interest, we adopted $\kappa=0.8$ for galaxies in the Fornax cluster, $\kappa=0.02$ for galaxies in the Perseus cluster, and $\kappa=12000$ for the NGC\,6744 field. NGC\,6744 is at the threshold for resolving individual stars, and a very high $\kappa$ value is needed to ensure that each star is not masked.

\subsection{\label{sec:sbf_mags}Extracting SBF fluctuation magnitudes}
The amplitude of the SBF signal is measured on the spatial power spectrum of the masked, background- and model-subtracted image in a single annulus, which is manually set in the configuration file. For massive galaxies, the inner radius is typically chosen to avoid the very centre of the galaxy, where the concentration of GC is highest and our ability to detect them is hampered by the increased luminosity of the galaxy. This is also the region where model residuals are often present. In less massive dwarf galaxies it is possible to use a very small inner radius; any GC or nuclei present in the central regions above our detection thresholds are easy to mask automatically, and fainter sources can be masked manually. For all galaxies, the outer radius of the annulus is set to avoid the low-S/N regions of the galaxy outskirts. 

A two dimensional (2D) power spectrum can be created from the Fourier transform $\hat{f}$ by taking $P = |\hat{f}|^2 $. From the power spectrum image, we extract an azimuthally averaged radial profile which can be fit with a function of the form  
\begin{equation}
\label{eq:power}
P(k) = P_\mathrm{0} \, E(k) + P_\mathrm{1},
\end{equation}
where $P_\mathrm{0}$ is the total amplitude of the fluctuations from stellar and spurious sources, $P_\mathrm{1}$ is the white-noise component -- expected to be flat at high wavenumbers -- and $E(k)$ is the expectation power spectrum. The units of these terms are the same units as the input image, where any any spatial term has been converted to wavenumbers; for the \Euclid ERO images, the images have been scaled into arbitrary ADU\,s$^{-1}$ flux units \citep{EROData}. The shape of $E(k)$ is obtained by convolving the power spectrum of the local, normalised PSFs with that of the window function (i.e., the cumulative mask that accounts for the SBF annulus, compact and extended objects, and manually masked sources and model residuals). We obtain independent $P(k)$ fits for each PSF in the sample, then extract the median $P_\mathrm{0}$ and $P_\mathrm{1}$ values from all fits as the representative values for the galaxy.

The interpolation kernel used to combine the individual images in the pipeline and the large-scale residual subtraction can corrupt the power spectrum, $P(k)$, at large and small wavenumbers, respectively (see Fig.\,\ref{fig:process_example1}). We avoid these regions by fitting the shape of power spectrum within a specified range of wavenumbers. In the present version of FAST-SBF, the lower and upper limits of this range, $k_{\mathrm{ini}}$ and $k_{\mathrm{end}}$, respectively, are determined empirically; the impact of these bounds will be discussed in Sect.\,\ref{sc:stability}. The quality of the power spectrum fit can be assessed in the subplots below the power spectrum in Fig.\,\ref{fig:process_example1}, where the middle panel shows the difference between the best fit and the data, and the lower panel shows the $P_\mathrm{0}$ values that are returned by iterating the fitting procedure, increasing $k_{\mathrm{ini}}$ by removing one additional data point in each iteration. 

The best-fit $P_\mathrm{0}$ value is obtained from the $P_\mathrm{0}$--$k_{\mathrm{ini}}$ plot by identifying a region with a flat profile, where $P_\mathrm{0}$ remains relatively constant. This interval is defined in the code as a fraction of the range over which $P(k)$ is fit, such that the relative number of data points used for every galaxy is consistent. Within this range, we take the median value as the best-fit value of the variance $P_\mathrm{0}$. Deviations at small and large $k$ values are expected, as they result from the processing stages described above.

The measured $P_\mathrm{0}$ value still includes contributions for the spurious sources, described in Sect.\,\ref{sec:spurious}. The spurious source term, $P_\mathrm{r}$, is subtracted from the total amplitude to leave the stellar SBF term: $P_ \mathrm{f} = P_\mathrm{0} - P_\mathrm{r}$. Once the stellar variance term, $P_\mathrm{f}$, is estimated, the apparent SBF magnitude $\overline{m}$ can be calculated from
\begin{equation}
\overline{m} = -2.5 \log_{10}\left(\frac{P_\mathrm{f}}{[\mathrm{ADU}\,\mathrm{s}^{-1}]}\right) + m_\mathrm{zp},
\end{equation}
\noindent where $m_\mathrm{zp}$ is the photometric zero point of the band used for the measurements.

\subsection{Measuring colours}
The colours that we report in this work are measured in the same annulus and using the same mask as the SBF signal. To account for the different pixel scales in the \IE and \HE bands, the \IE mask is projected onto the \HE image using the {\texttt{Astropy reproject\_exact}} function \citep{Robitaille2020}. Fluxes are extracted from the sky-background subtracted, masked images, and then converted to magnitudes to obtain a colour. 

Given the wide wavelength separation between the \IE and \HE filters, it is possible that there will be spurious sources, particularly background galaxies, in the \HE band that are not detected in the \IE image. We produce colour maps and radial colour profiles to visually check if this is an issue. Only two of the massive Fornax cluster members showed anomalously red sources with no \IE counterpart in the SBF annulus. However, the sources were relatively small, and had a minimal impact on our measurements. After re-running the procedure and manually masking these objects, the changes in  $\overline{m}$ and colour were negligible, well within the photometric errors estimated for both quantities. 

The uncorrected colours are reported in Table\,\ref{table:results}, along with the extinction corrections. We use the extinction corrected values for the calibration in Sect.\,\ref{sc:calibration}. {\texttt{FAST-SBF}} queries the IRSA dust maps \citep{Schlafly2011} using  the {\texttt{IRSA Galactic Dust Reddening and Extinction Query Tool}} from {\texttt{Astropy}} to obtain $E(B-V)$ at the location of the galaxy. We use $A_{\IE} = 2.12\, E(B-V)$ and $A_{\HE} = 0.47\, E(B-V)$, adopting the coefficients calculated by \citet{EROPerseusDGs}. An error of 10\% is assumed for the extinction corrections for both SBF magnitudes and colours.

\begin{figure*}[tbp]
  \begin{center}
    \includegraphics[width=1.98\columnwidth]{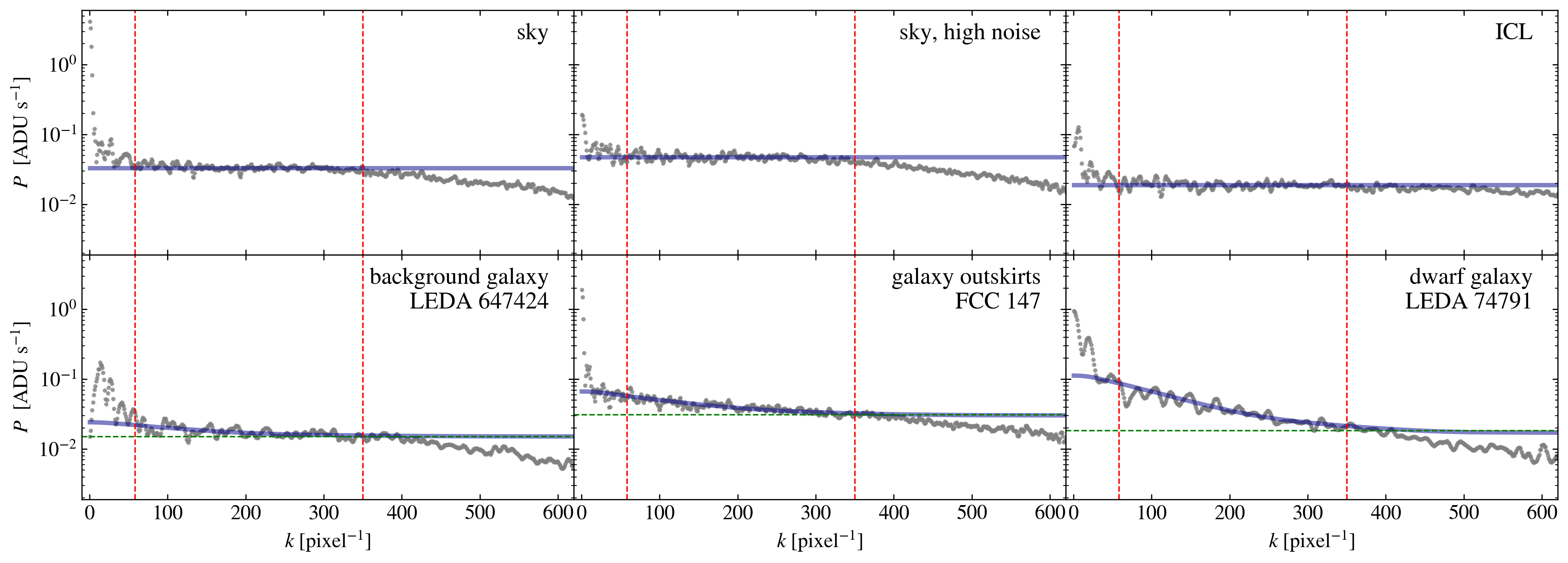}
   \end{center}
\caption{\emph{Top row}: the power spectrum of three sky regions across the Fornax ERO image: an area with a typical noise level (\emph{left}), in a stripe with higher noise (\emph{middle}), and in the region with ICL (\emph{right}). \emph{Bottom row, from left to right}: the power spectrum of the background galaxy LEDA\,647424 ($d{\sim} 250$\,Mpc; \citealt{Morris2007}), in the halo of the massive cluster member FCC\,147, and for the dwarf cluster member LEDA\,74791. The data (grey points) are fit with a function of the form $P(k) = P_\mathrm{0} \, E(k) + P_\mathrm{1}$ (blue line; see text for details) in the $k$ interval between the red dashed lines. The extracted white noise component, $P_\mathrm{1}$, is plotted as a dashed green line when it can be distinguished from the best-fit line.}
\label{fig:sky}
\end{figure*}

\subsection{\label{sec:SBFsum}FAST-SBF output}

The \texttt{FAST-SBF} code returns $\overline{m}$ values, colours, and a number of data products that can be used to test the robustness of the measurements at various stages of the code. The most relevant parameters are  listed in Table \ref{table:results}.
In the following section, we test the robustness of the measurements in \Euclid data, derive the $\overline{M}_\mathrm{IE}$ calibration for the Fornax cluster, and test the calibration by applying it to the two other test fields.

\section{\label{sc:results}Results}

We applied the {\texttt{FAST-SBF}} code to 31 galaxies in the Fornax ERO data. Notes on the fits of select galaxies are included in Appendix\,\ref{appendix:individual}. In the following sections, we will focus on the azimuthally averaged radial profile of the power spectrum to test the stability and reliability of our measurements.

\subsection{\label{sc:SBFsky}Power spectrum of the sky background}

Before running the pipeline on the galaxy sample, we first applied {\texttt{FAST-SBF}} to empty regions of the sky. Such areas should only contain fluctuations due to faint, unmasked background galaxies; GC are not expected at large distances from their host galaxies and prior studies indicate that Fornax has a small fraction of intra-cluster GC \citep{Cantiello2020}. This serves as a test to verify if, and how, the specific ERO data reduction strategy affected the power spectrum of the background. For this test, we adopted the same $r_\mathrm{in}$ and $r_\mathrm{out}$ values as the annulus of one of the largest dwarf candidates in the field, FCC\,188; the dwarf is large enough that the annulus has good statistics (${>}10^4$\,pixels) but small enough that relatively clean patches of sky of this size can still be found. Any visible sources remaining in the annulus were either masked using {\texttt{Source Extractor}} catalogues (bright sources) or by hand (faint sources).

This test was run multiple times: in a clean region of sky with a typical noise level ($\mathrm{RA}=\ra{3;35;1.31}$, $\mathrm{Dec}=\ang{-35;1;18.67}$), in a strip of sky with visibly higher noise ($\mathrm{RA}=\ra{3;35;28.78}$, $\mathrm{Dec}=\ang{-35;11;57.45}$), and in the area where the ICL has been observed (in the vicinity of FCC\,188, at $\mathrm{RA}=\ra{3;36;44.02}$ and $\mathrm{Dec}=\ang{-35;37;9.45}$). The resulting power spectra are shown in Fig.\,\ref{fig:sky}. In all three cases, the power spectrum in the region of interest is distinctly flat over the wavenumber range used for the power spectrum fitting (between the red-dashed vertical lines in the figure). The slight uptick observed at low $k$ values is likely due to the residual spurious sources, as discussed in Sect.\,\ref{sec:spurious}. The flat power spectrum in all three cases confirms that the data reduction pipeline has not introduced any correlations that would mimic an SBF signal and that the ICL will not affect the power spectrum of any galaxies in that region of the cluster. As a proof of concept, we also note that the measurement in the region with higher noise has the highest white noise component, as expected.

We also extend this test to three additional sources in the Fornax ERO image, seen in the bottom row of Fig.\,\ref{fig:sky}: the background galaxy LEDA\,67424 ($cz\,{=}\,\num{19057}$\,km\,s$^{-1}$; \citealt{Morris2007}), the halo of a massive Fornax cluster member, FCC\,147, and the dwarf cluster member LEDA\,74791. The background galaxy was selected because no stellar fluctuation signal is expected at that distance with the available data; given the measured velocity and assuming a Hubble constant $H_0 = 74$\,km\,s$^{-1}$\,Mpc$^{-1}$ \citep{Jensen2025,Casertano2025,Garnavich2023}, one would expect $\overline{m}_\mathrm{IE}\sim36$. The test in the halo of FCC\,147 was done at a galactocentric distance of \ang{;2.5;} ($\sim6\,R_\mathrm{e}$), in close proximity to the dwarf galaxy LEDA\,74747, in order to verify the presence of stellar fluctuation contamination from the massive galaxy in the dwarf power spectrum. Finally, the galaxy LEDA\,74791, a relatively isolated dwarf, is included as a reference. For the two LEDA galaxies, we used an annulus appropriate for the size of the galaxy, while in the halo of FCC\,147 we used the same-sized annulus as the tests of the sky.

The spectrum of the background galaxy is relatively flat, similar to that of the measured sky regions, but some residual fluctuation term remains. This arises from the undetected and unmasked GC in the targeted galaxy, which are the main source of contamination to the stellar fluctuation signal when the photometry is not deep enough to properly sample and mask them \citep{Blakeslee1997}. At the distance of LEDA\,67424, the GCLF is expected to have a turnover magnitude $m_{\mathrm{TOM}}\sim27.5$ in \IE, which is fainter than \Euclid's 5$\sigma$ point-source depth \citep{EROData}. The fact that we measure a $P_\mathrm{r}$ correction term that is greater than the amplitude of the stellar fluctuations further supports this interpretation. Constraining the maximum distance we can reach with SBF in \Euclid data will be the subject of a future work.

The contributions from unmasked GC and background galaxies are expected to be minimal in the halo of FCC\,147, given the size of the annulus and its distance from the galaxy photocentre. Indeed, the spurious source correction term is only one tenth of the total fluctuation amplitude, $P_\mathrm{0}$, indicating that we are measuring a stellar fluctuation signal even at such large distances from the galaxy centre. While this is promising for measuring the signal of the massive galaxy even at large radii, this also means that the signal and colour of the dwarf galaxy in the halo, LEDA\,74747, are contaminated. Several other dwarfs are similarly affected, and will be removed from the calibration analysis in Sect.\,\ref{sc:calibration}.

\subsection{\label{sc:stability}Stability of fits}
One of the challenges of measuring SBF with the \Euclid dataset is extracting $\overline{m}_\mathrm{IE}$ from the power spectrum of the residual masked frame. Under ideal conditions, the power spectrum has a well-defined plateau at high wavenumbers, which helps constrain the global power spectrum fit and measure the white noise component. As seen in Figs.\,\ref{fig:process_example1} and \ref{fig:sky}, however, the plateau is not very prominent in the \IE data. Such behaviour has already been observed in previous datasets \citep{Cantiello2018,Cantiello2005,Mei2005IV}, but the absence of a clear white-noise plateau appears to be even more severe in the \Euclid ERO data. 

This lack of a plateau in the power spectrum arises primarily from the stacking and geometric correction approaches used in the ERO and standard \Euclid reduction pipelines. A key component of our ongoing work is the development of a dedicated data-processing pipeline for galaxies in the nearby Universe, which can mitigate the issue. Here, we test whether \texttt{FAST-SBF} can still converge to stable fluctuation amplitudes across a range of input parameters and fitting strategies, despite the weak plateau.

For the subsample of nine galaxies that are confirmed cluster members based on prior SBF measurements, we initially ran the code, fine tuning the parameters in {\texttt{FAST-SBF}} until a set of optimal results were obtained. We consider the output of {\texttt{FAST-SBF}} to be of good quality when all of the intermediate data products (e.g., the model residuals, masks, and the power spectrum fits) are satisfactory, as determined by a visual inspection of the images and plots. Using the parameters from the optimal run as a `reference run', we then re-ran the code multiple times, varying a single parameter each time to test the stability of our results. We tested: (a) the window interval within which we fit the power spectrum, (b) $\kappa$, the parameter that controls the detection rate of bright sources on top of the galaxy, and (c) the annular region from which the SBF measurement is taken. 

\begin{figure*}[tbp!]
  \begin{center}
    \includegraphics[width=1.95\columnwidth]{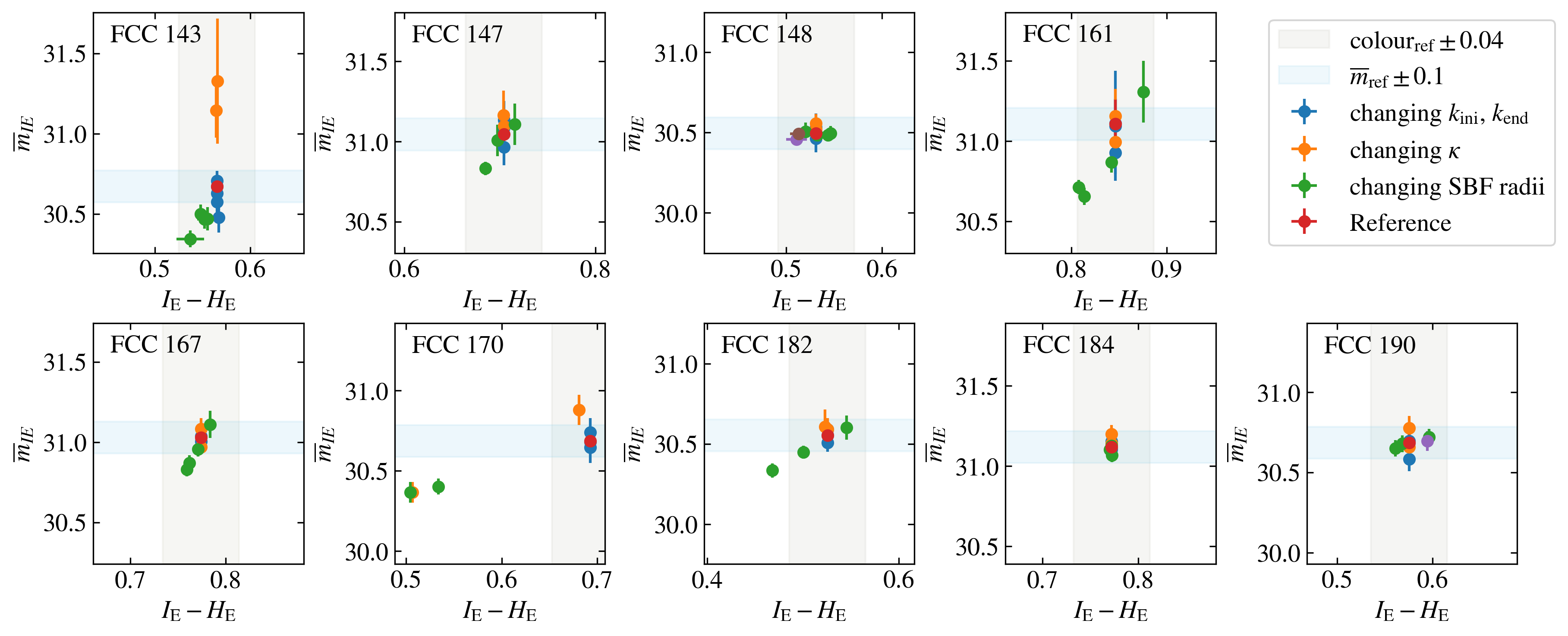}
   \end{center}
\caption{Stability of the {\texttt{FAST-SBF}} results. For the nine galaxies with prior SBF measurements, we first obtained a good (or reasonable, in the case of FCC\,170) galaxy model with a flat residual, then varied the input parameters to test the stability of our results. We changed: (a) $\kappa$, the deblending parameter that governs how point sources are detected above the surface brightness of the galaxy (orange); (b) the fitting interval used for the modelling the power spectrum ($k_\mathrm{ini}$ and $k_\mathrm{end}$; blue); and (c) the annulus in which the SBF signal is extracted (green). For each galaxy, we also included shaded bands around the reference value to highlight the expected errors: $\overline{m} \pm 0.1$ and colour $\pm0.04$. The output of the `reference run' for each galaxy is shown in red.}
\label{fig:fits_stability}
\end{figure*}

The results of these runs are shown in Fig.\,\ref{fig:fits_stability}. To help guide the eye, we also include shaded regions that correspond to the level of error we expect in the final output: $\pm0.1$ in $\overline{m}_\mathrm{IE}$ and $\pm0.04$ in colour \citep[e.g.,][]{Cantiello2024}. With few exceptions, the values returned by each run are well within the expected errors. The largest scatter is found when the size of the annulus is changed, but this is expected because concentric annuli will probe different underlying stellar populations in the host galaxy; this will be discussed in more detail in Sect.\,\ref{sc:sspmodels}. 

Changing $\kappa$ can also lead to errors that are on par with, or slightly larger than, the expected error in $\overline{m}_\mathrm{IE}$. However, a visual inspection quickly reveals that moderate offsets in $\kappa$ result in either excessive or insufficient masking of contaminating GC and background galaxies, which directly impacts the measured $P_\mathrm{f}$ and  $\overline{m}$. We therefore adopt an empirically determined optimal value of $\kappa$ based on the full sample of galaxies in a given field, and keep it fixed thereafter. Note that the code has a non-linear response to $\kappa$, which is why the scatter from this parameter is not symmetric.

We also tested the stability of the $P_\mathrm{0}$ and $P_\mathrm{1}$ values extracted from the global fit by varying the wavenumber fitting interval. As long as a reasonable range is selected, there is little impact on the resulting measured $\overline{m}_\mathrm{IE}$ values. Poor fits at more extreme values can be visually flagged from the plot of the power spectrum itself, or the subplots of Fig.\,\ref{fig:process_example1} that show (a) the difference between the data and best fit and (b) the stability of $P_\mathrm{0}$ as a function of different $k$ intervals.

\begin{figure}[tbp!]
  \begin{center}
    \includegraphics[width=1.\columnwidth]{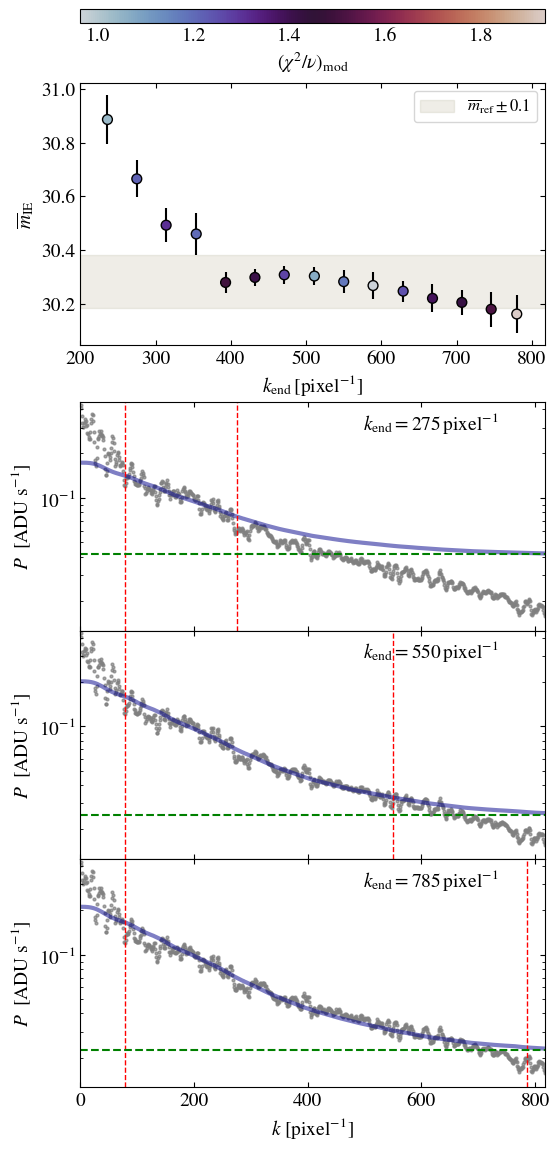}
   \end{center}
\caption{Testing the sensitivity of our fitting routine on $k_\mathrm{end}$ for a single galaxy: FCC\,188.  \emph{Top}: the $\overline{m}_\mathrm{IE}$ we extract as a function of $k_\mathrm{end}$, when all other parameters are held constant. The $(\chi^2/\nu)_\mathrm{mod}$ values are a modified reduced chi-squared statistic; see the text for a discussion. \emph{Lower three panels}: the power spectrum of the galaxy for three values of $k_\mathrm{end}$. The colours are the same as in Fig.\,\ref{fig:sky}. }
\label{fig:k_end}
\end{figure}

Based on the appearance of the power spectrum, one might expect that $k_{\mathrm{end}}$, the upper limit of the fitting range, would have a stronger impact on our ability to fit the white-noise plateau than $k_{\mathrm{ini}}$. We explicitly test this, and further explore the stability of the global fit, for a single galaxy in the sample: FCC\,188. Holding $k_\mathrm{ini}$ and all other parameters fixed, we re-ran the code multiple times, increasing $k_\mathrm{end}$ with each iteration.\footnote{{\texttt{FAST-SBF}} defines the interval within which the total variance $P_\mathrm{0}$ is measured as a fraction of the window size used for the global fit to the data (see Sect. \ref{sec:sbf_mags}). This will also increase for a fixed $k_\mathrm{ini}$ and increasing $k_\mathrm{end}$, but is not expected to impact the results of this test.} The results, seen in Fig.\,\ref{fig:k_end}, suggest that the output is not particularly sensitive to $k_\mathrm{end}$, and is fairly stable as long as $k_\mathrm{end}$ reaches wavenumbers $k\geq400\,\mathrm{pixel^{-1}}$. It is important to note that optimal $k_\mathrm{ini}$ and $k_\mathrm{end}$ values scale with the size of the cutout used for the measurements (which may be increased to allow for the detection of more PSF candidates, or decreased for small galaxies with correspondingly small annuli to reduce computational time), so the condition that $k_\mathrm{end} > 400 \,\mathrm{pixel^{-1}}$ should not be applied blindly. In particular, the examples in Fig.\,\ref{fig:sky} were run on a smaller cutout for faster runtimes, and the fitting window was scaled accordingly. 

In the current version of \texttt{FAST-SBF}, we have not yet implemented a formal metric to quantify the goodness of fit for the global power spectrum. For this test, we estimated 
\begin{equation}
\chi^2=\sum_i \frac{[y_i - f(x_i)]^2}{\sigma_i^2}\;,
\end{equation}
where $y_i$ are the measured $P(k)$ values, $f(x) = P_\mathrm{0} \, E(k) + P_\mathrm{1}$ is the best fit to the data, and $\sigma_i$ is the root mean square (RMS) derived from the azimuthal average of the $i$-th annulus. We do not report the reduced chi-squared statistic, $\chi^2/\nu$, because the RMS values are larger than the residuals, $y_i - f(x_i)$, at any given data point, leading to $\chi^2/\nu \ll 1$. However, we can account for sample size and approximate $\chi^2/\nu$ as $(\chi^2/{\nu)_\mathrm{mod}} =  100\,\chi^2/ (k_\mathrm{end} - k_\mathrm{ini})$. Although this scaling is arbitrary, the relative values are still illustrative.  

As seen in Fig.\,\ref{fig:k_end}, the $(\chi^2/{\nu)_\mathrm{mod}}$ values reach a minimum at $k_\mathrm{end} \sim 600 \,\mathrm{pixel^{-1}}$, in the middle of the $\overline{m}_\mathrm{IE}$ plateau. The global power spectrum, shown for $k_\mathrm{end} = 550 \,\mathrm{pixel^{-1}}$ in the lower panels of the figure, also demonstrates that this $k_\mathrm{end}$ returns a reasonable fit. In contrast, at small $k_{\mathrm{end}}$ values, the global fit is poor and the extracted $\overline{m}_\mathrm{IE}$ values are unstable, despite the relatively low $(\chi^2/{\nu)_\mathrm{mod}}$ values. At high $k_{\mathrm{end}}$, the combination of  a poorer fit to the global power spectrum and increasing $(\chi^2/{\nu)_\mathrm{mod}}$  values leads us to reject $k_{\mathrm{end}}$ values above $650\,\mathrm{pixel}^{-1}$.

Overall, the fitting routine appears to be robust for 18 galaxies, the eight massive galaxies with prior SBF measurements and 10 dwarfs. Dwarf galaxies are expected to have a larger scatter around the $\overline{m}$ versus colour relation than the massive galaxies, by roughly a factor of two or more \citep[e.g.,][]{Cantiello2024,Jensen2015,Blakeslee2009}. However, the fits were often found to be unstable for the smallest and faintest dwarfs in the sample; small changes in the SBF annulus or the masking routine could generate deviations in $\overline{m}_\mathrm{IE}$ far in excess of the estimated error, with no clear reason to prefer the output from one run over the other. For the dwarf galaxies with unstable results, we consider our measurements solely as confirmation of group membership, as detailed in Sect. \ref{sc:calibration}.

\subsection{\label{sc:calibration}Deriving the zero point of the $\overline{M}_\mathrm{IE,0}$ versus $(\IE{-}\HE)_0$ relation}

\begin{figure}[tbp!]
  \begin{center}
    \includegraphics[width=0.99\columnwidth]{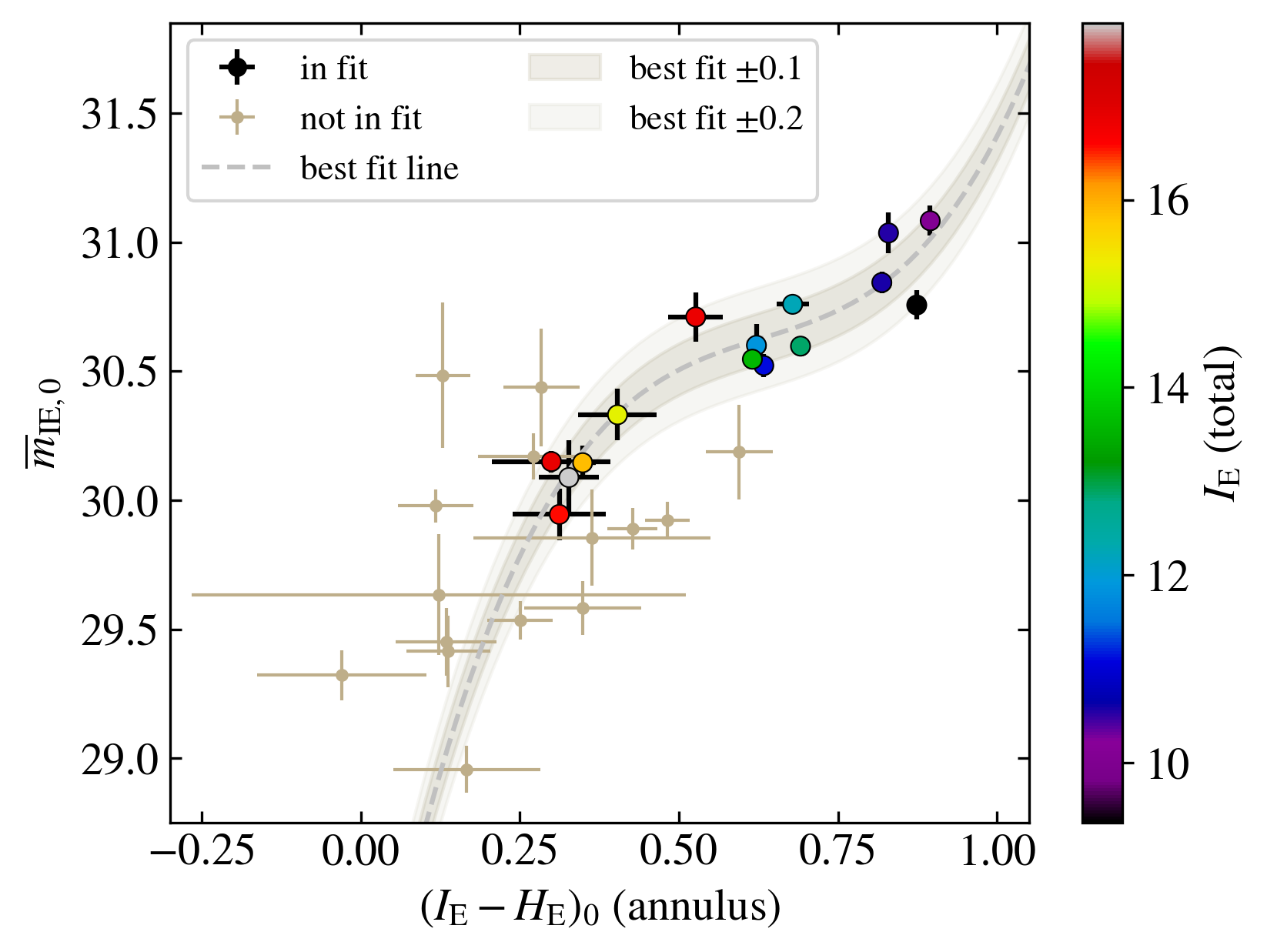}
   \end{center}
\caption{Colour-magnitude diagram for the SBF targets in the Fornax galaxy cluster. The colour and SBF magnitude $\overline{m}_\mathrm{IE,0}$ are measured in the same annulus, and the galaxies used for the best-fit line are colour coded according to the total apparent magnitude that we measure for the galaxy. Around the best-fit line, we include two shaded regions, denoting uncertainties of $\overline{m}_\mathrm{IE}\pm0.1$ and $\overline{m}_\mathrm{IE}\pm0.2$ from the best-fit line; these are the typical uncertainties that we would expect from this process for massive and dwarf galaxies, respectively. For illustration purposes, the galaxies that were not included in the fit are shown in tan; see text for details. }
\label{fig:color-mbar}
\end{figure}

To extract distances from SBF, we must first calibrate the stellar population dependence of the $\overline{M}_\mathrm{IE,0}$ versus $(\IE{-}\HE)_0$ relation and derive the zero point. To ensure that this calibration is not skewed by outliers, we restrict the sample to the Fornax ERO galaxies with the most robust measurements. We first applied a selection cut on the total magnitude of the galaxy, keeping the galaxies with $\IE < 18.5$. Fainter sources (in this sample) typically suffer from two problems: (i) they are small with correspondingly few  pixels for SBF measurements, making the measurements generally unstable, and (ii) they typically have very low surface brightnesses, which makes them more susceptible to contamination (e.g., from neighbouring sources or poorly estimated backgrounds). We also removed six dwarfs from the sample, whose proximity to a massive neighbour suggests that their colours, and possibly their SBF magnitudes, are contaminated. Basic properties of the 15 galaxies (massive and dwarf) that remain in the sample can be found in Table\,\ref{table:results}.

In Fig.\,\ref{fig:color-mbar}, we plot the extinction corrected $\overline{m}_\mathrm{IE,0}$ versus $(\IE{-}\HE)_0$ colour measurements for all galaxies in the sample. The 15 galaxies that remain in the calibration sample are also plotted as a function of the total \IE magnitude we measure for the galaxy, while the candidates that will not be used in the calibration are included to show the scatter in the measurements. There is a clear trend, which we fit with a third-order polynomial, following previous works \citep[e.g.,][]{Cantiello2024}. 

We set the zero point of the method in the \IE filter using the distance modulus for Fornax, $(m-M) = 31.42$ \citep{Anand2024}. Based on our best-fit line, the absolute SBF amplitude $\overline{M}_\mathrm{IE,0}$ at a fixed galaxy colour $(\IE{-}\HE)_0$ is then derived as
\begin{equation}
\label{eq:fit}
\overline{M}_\mathrm{IE,0} = - 0.78 + 0.76\,x + 0.66\,x^2  + 9.87\,x^3 ,
\end{equation}
\noindent where $x = (\IE{-}\HE)_0 - 0.65$. Following conventions in the literature, we give the best-fit line as a function of the colour with respect to a reference colour $x-x_\mathrm{ref}$. When $x_\mathrm{ref}$ is near the sample median, this format allows for quick estimations of the distance modulus. Here we adopt $x_\mathrm{ref}=0.65$.

The massive galaxies tightly follow the best-fit line, with a typical scatter $\Delta \overline{m} < 0.1$. The dwarfs have a larger scatter, as expected \citep[e.g.,][]{Cantiello2024,Jensen2015,Blakeslee2009}, but still fall within $\Delta \overline{m} < 0.2$, although it should again be emphasized that these are the dwarf galaxies with the most robust measurements. The dwarf galaxies not included in the fit are scattered around the projection of the best fit line towards bluer colours, but this must be interpreted with caution, as the steepness of the slope is driven by a single data point. The blue side of the curve can be further characterised with more measurements from forthcoming data releases.

Although we manually excluded six dwarf candidates in the halos of massive galaxies from this analysis, they are likely also cluster members. Some of the galaxies have spectroscopic measurements in the literature confirming cluster membership (e.g., LEDA\,74747, discussed in Sec.\,\ref{sc:SBFsky}; \citealt{Dabringhausen2016}), but contamination from the neighbouring galaxies has resulted in $\overline{m}$ measurements that are too bright, colours that are too red, or a combination of the two problems, effectively shifting the galaxies away from the colour-magnitude relation that we find for the most robust cluster members. Nevertheless, the strong SBF signal that we detect supports their inclusion as cluster members, and we use this argument to tentatively confirm dwarf candidates without available spectroscopic data (e.g., FDS\,110299) as cluster members. We also tentatively classify the fainter dwarfs, those for which we detect a fluctuation signal but obtain unstable results, as group members. The very fact that we can measure a stellar fluctuation signal at all, and one that scatters around the expected $\overline{m}_\mathrm{IE,0}$ for the measured colours, suggests that these galaxies are also cluster members. The properties of the likely cluster members are given in Appendix\,\ref{appendix:tables} (Table\,\ref{table:maybe}), and all excluded galaxies are shown in Fig.\,\ref{fig:color-mbar} as light brown points.

\begin{table*}[tbp!]
\newcommand{\pd}{\phantom{1}}
\setlength{\tabcolsep}{3.25pt}
  \caption{Key parameters of the galaxies in the Fornax cluster used to standardise the SBF distance zeropoint (the coloured points in Fig.\,\ref{fig:color-mbar}). For a given id, \texttt{FAST-SBF} queries databases for the coordinates (given in degrees) in the following order: (a) NED, (b) SIMBAD \citep{Wenger2000}, or (c) a user input file.  For the dwarfs galaxies, preference was given to the coordinates from the {\texttt{JAFAR}} annotation tool, which were input manually. The T-Type morphologies are from HyperLeda \citep{Makarov2014}. The remaining values are output from {\texttt{FAST-SBF}}: the effective radius $R_\mathrm{e}$ in arcseconds, the ($\IE{-}\HE$) colour, apparent SBF magnitude $\overline{m}$, the extinction correction, inner $r_\mathrm{in}$ and outer $r_\mathrm{out}$ radii of the annulus for the SBF measurement, and the number of unmasked pixels $N_\mathrm{pix}$.} 
\smallskip
\label{table:results}
\centering
\smallskip
\small
\begin{tabular}{l l c c c r c c c r r }
  \hline
  & & & & & & & & &\\[-7pt]
\omit\hfil Name\hfil &\omit\hfil Alt Name\hfil  & \omit\hfil RA \hfil & \omit\hfil Dec\hfil & \omit\hfil $T_\mathrm{type}$ \hfill &\omit\hfil $R_\mathrm{e}$\hfil  &\omit\hfil $(\IE{-}\HE)$\hfil & \omit\hfil $\overline{m}_\mathrm{IE}$ \hfil &  \omit\hfil $E(B-V)$\hfil & \omit\hfil $r_\mathrm{in}/r_\mathrm{out}$\hfil & \omit\hfil $N_\mathrm{pix}$\hfil  \\
 & & \omit\hfil[deg] \hfil &\omit\hfil\,[deg] \hfil& & \omit\hfil [arcsec] \hfil & & & & \omit\hfil [pixel] \hfil & \omit\hfil[pixel] \hfil \\
  & & & & & & & & &\\[-8pt]
\hline
  & & & & & & & & &\\[-8pt]
FCC\,133 & LEDA\,74737 & 53.584124 & $-$35.362470 &  $-3.5\pm2.1$  & 7.85 & 0.547$\pm$0.043 & 30.74$\pm$0.09 & 0.0125 &  12/50$\phantom{0}$ & 6544 \\ 
FCC\,143 & NGC\,1373 & 53.746690 & $-$35.171070 & $-3.6\pm1.7$ & 6.93 & 0.711$\pm$0.005 & 30.62$\pm$0.04 & 0.0120 &  32/150 & 62734 \\ 
FCC\,147 & NGC\,1374 & 53.819160 & $-$35.226270 & $-4.5\pm1.0$  & 18.36 & 0.839$\pm$0.001 & 30.87$\pm$0.04 & 0.0119 &  16/256 & 189183 \\ 
FCC\,148 & NGC\,1375 & 53.820030 & $-$35.265634 & $-2.2\pm0.8$  & 26.30 & 0.642$\pm$0.010 & 30.63$\pm$0.08 & 0.0120 &  64/256 & 83727 \\ 
FCC\,156 & LEDA\,13287 & 53.928090 & $-$35.338080 & $-3.5\pm2.1$  & 8.63 & 0.319$\pm$0.093 & 30.17$\pm$0.04 & 0.0118 &  2/76$\phantom{0}$ & 17213 \\ 
FCC\,161 & NGC\,1379 & 54.016503 & $-$35.441251 & $-4.8\pm0.5$& 20.41 & 0.846$\pm$0.001 & 31.06$\pm$0.08 & 0.0102 &  16/200 & 116783 \\ 
FCC\,167 & NGC\,1380 & 54.114876 & $-$34.976034 & $-2.3\pm0.7$  & 29.56 & 0.898$\pm$0.001 & 30.79$\pm$0.06 & 0.0145 &  100/350 & 317275 \\ 
FCC\,170 & NGC\,1381 & 54.131910 & $-$35.295140 &  $-2.1\pm0.7$ & 13.47 & 0.652$\pm$0.014 & 30.55$\pm$0.05 & 0.0114 &  128/256 & 70324 \\ 
FCC\,175 & LEDA\,74775 & 54.179117 & $-$35.435581 & $-3.5\pm 2.1$ & 9.46 & 0.345$\pm$0.047 & 30.11$\pm$0.15 & 0.0108 &  4/50$\phantom{0}$ & 6958 \\ 
FCC\,181 & LEDA\,74781 & 54.221908 & $-$34.938466 & $-4.3 \pm 1.6$ & 7.25 & 0.337$\pm$0.073 & 29.98$\pm$0.10 & 0.0149 &  10/64$\phantom{0}$ & 11931 \\ 
FCC\,182 & $\cdots$ & 54.226340 & $-$35.374660 & $-2.8\pm0.6$  & 8.32 & 0.634$\pm$0.009 & 30.57$\pm$0.03 & 0.0113 &  32/256 & 191963 \\ 
FCC\,184 & NGC\,1387 & 54.237640 & $-$35.506530 & $-2.8\pm0.5$ & 17.02 & 0.913$\pm$0.015 & 31.11$\pm$0.06 & 0.0109 &  150/350 & 298645 \\ 
FCC\,188 & $\cdots$ & 54.268730 & $-$35.590410 & $-4.3\pm1.6$ & 10.99 & 0.422$\pm$0.062 & 30.35$\pm$0.10 & 0.0112 &  4/86$\phantom{0}$ & 22916 \\ 
FCC\,190 & NGC\,1382b & 54.287384 & $-$35.194999 & $-2.8\pm0.6$ & 16.12 & 0.702$\pm$0.025 & 30.79$\pm$0.04 & 0.0140 &  64/256 & 140979 \\ 
FCC\,195 & LEDA\,74791  & 54.347517 & $-$34.899969 & $-3.5\pm2.1$ & 10.47 & 0.370$\pm$0.021 & 30.17$\pm$0.07 & 0.0128 &  5/50$\phantom{0}$ & 7745 \\ 

\hline
\end{tabular}
\end{table*}

\subsection{\label{sc:other_distances}Testing the SBF signal measurements at other distances}
\begin{figure}[tbp!]
  \begin{center}
    \includegraphics[width=.98\columnwidth]{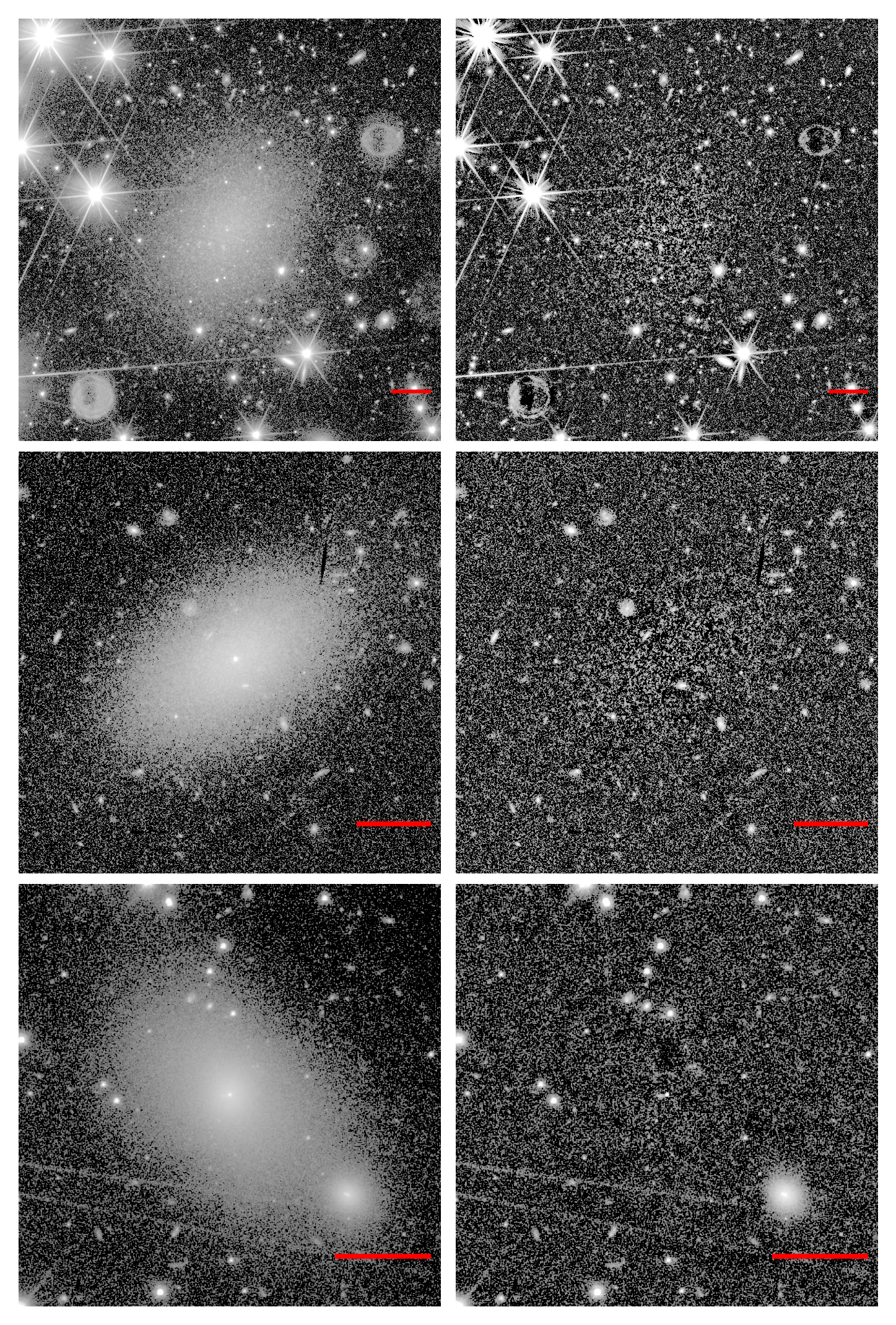}
   \end{center}
\caption{Background subtracted (\textit{left}) and background and model subtracted (\textit{right}) \IE images of three galaxies from this work. \emph{From top to bottom}, the galaxies are ordered by increasing distance: dw1906m6357 ($d\sim9.6$\,Mpc), LEDA\,74781 ($d\sim19.3$\,Mpc), and edwc0331 ($d\sim69.5$\,Mpc). The two nearest galaxies have similar colours, while edwc0331 is the bluest galaxy among the Perseus cluster members we tested, thus the amplitude of the fluctuations correlates closely with the distance. The red line on each cutout marks \ang{;;10}. }
\label{fig:SBF_signal_compare}
\end{figure}

Having standardized the relation in \IE, we can estimate distances to other systems in the \Euclid ERO images, and compare our values against previous distance estimates to test the calibration. Three example images, one from the Fornax cluster, the NGC\,6744 system of dwarf satellites, and one from Perseus, are shown in Fig.\,\ref{fig:SBF_signal_compare}, to illustrate the impact of distance on the SBF amplitudes.

\subsubsection{\label{sc:Fornax}Fornax cluster}

The most direct comparison for our measurements is with existing SBF distances to Fornax cluster galaxies. To avoid differences between various studies, we restrict this comparison to the most recent SBF measurements by \citet{Blakeslee2009}. Furthermore, we explicitly exclude TRGB distances, since the zero point of our calibration equation is derived from the TRGB method. By matching the distance moduli for the eight galaxies in our sample that overlap with the \citet{Blakeslee2009} catalogue, we find a median difference in distance moduli of 0.06. Once we account for the systematic differences between the distance modulus of the cluster in the two works, however, this reduces to 0.02.

\subsubsection{\label{sc:Perseus}Perseus cluster}

\begin{figure}[tbp!]
  \begin{center}
    \includegraphics[width=0.95\columnwidth]{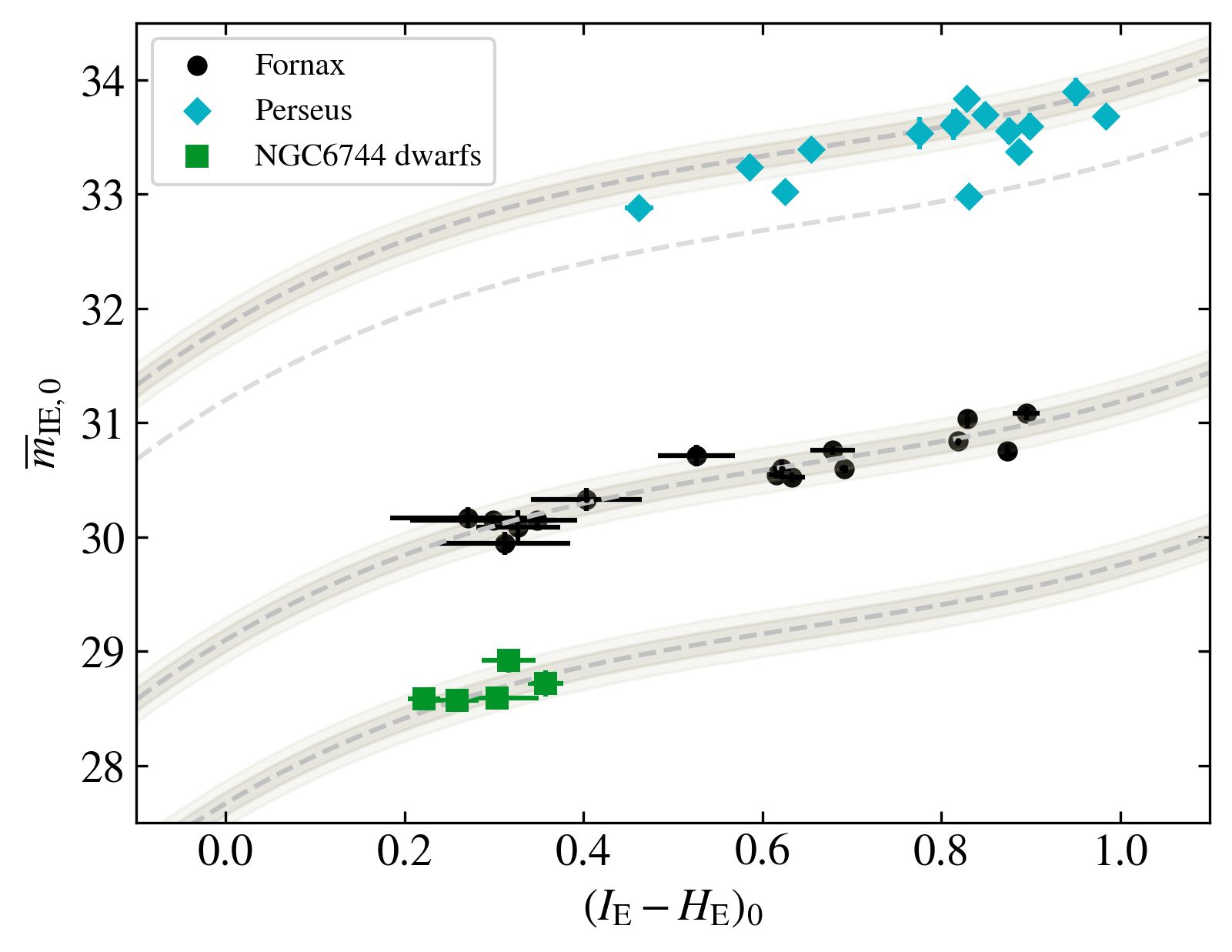}
   \end{center}
\caption{Colour-magnitude diagram comparing SBF measurements of galaxies in the Fornax cluster, Perseus cluster, and around NGC\,6744. The SBF calibration determined for the Fornax cluster galaxies (grey dashed line with the anticipated $\pm0.1$ errors shown as a shaded region; see Fig\,\ref{fig:color-mbar}) is plotted. This best-fit line and shaded region have also been shifted by the median difference of the galaxies in the other two systems, respectively, and replotted (see Sect.\,\ref{sc:other_distances} for details). We further highlight the galaxy suspected of being in the foreground of the Perseus cluster, by plotting the shifted best-fit line that gives its distance. }
\label{fig:color-mbar-multigroup}
\end{figure}

Of the 19 targets selected from the Perseus cluster catalogue, we were able to obtain reliable measurements for 15 galaxies; the other four galaxies had poor power spectrum fits, and were dropped from the subsequent analysis. Like the Fornax cluster sample, the remaining galaxies could be split into two categories: those were every internal check indicated that the procedure was robust (seven galaxies), and those for which one or more steps yielded reasonable -- but not optimal -- outputs (eight galaxies). All of the galaxies are shown in Fig.\,\ref{fig:color-mbar-multigroup}, but for this quick test we only use the seven most confident values to measure the relative distance. Taking the median distance between these seven galaxies and our best-fit line for Fornax, we measure $d_\mathrm{Perseus}/d_\mathrm{Fornax} = 3.6^{+0.4}_{-0.3}$. Having assumed the distance to Fornax,  $d_\mathrm{Fornax}=19.3$ Mpc (Sect. \ref{sc:calibration}), this implies $d_\mathrm{Perseus} = 69.5^{+4.6}_{-4.4}$\,Mpc, which is consistent with the aggregate distance found by \citet{Tully2009}: $(72\pm\,3)\,\mathrm{Mpc}$.

Interestingly, there is one galaxy among the seven good candidates, 2MASX J03171358+4126082, that appears to be a foreground object rather than a cluster member. Our measurements place it at a distance $d\,{\sim}\,50$\,Mpc. However, the galaxy has a spectroscopic redshift from SDSS ($z=0.01664$, or $v=\num{4990}$\,km\,s$^{-1}$; \citealt{Alam2015}) that is consistent with cluster membership ($v_\mathrm{Perseus} = \num{5006}$\,km\,s$^{-1}$; \citealt{Kang2024}). These two measurements can be reconciled if the galaxy is a foreground object that is being attracted towards the cluster with a relatively high velocity. The recessional velocity of a galaxy at 50\,Mpc due to the Hubble Flow alone (assuming $H_0 = 74$\,km\,s$^{-1}$\,Mpc$^{-1}$) is $v=\num{3700}$\,km\,s$^{-1}$, which implies that the additional velocity along the line of sight is approximately \num{1290}\,km\,s$^{-1}$. Such high line-of-sight velocities have been found for galaxies falling into massive clusters in simulations \citep{Borrow2023}. This galaxy illustrates the importance of redshift-independent distance estimates in mapping cluster populations. A more comprehensive study of the SBF amplitudes in galaxies within the Perseus ERO footprint is currently in progress.

\subsubsection{\label{sc:NGC6744}Testing the SBF signal around NGC 6744}

The nearly face-on SAB(r)bc galaxy NGC\,6744 is not a good target for SBF measurements, but we were able to obtain results for five dwarf galaxies in the ERO image \citep{ERONearbyGals}. The membership of the dwarfs was previously studied using SBF distances by \citet{Carlsten2022}, who concluded that four of the dwarfs were satellites, while the fifth (dw1909m6341) was listed as a potential satellite. 

Our measurements for these galaxies can be seen in Fig.\,\ref{fig:color-mbar-multigroup}, where they are tightly clustered. Using the median offset from the best-fit line for Fornax, we estimate the distance of the system to be $d_\mathrm{NGC6744}/d_\mathrm{Fornax} \,{=}\, 0.5\pm0.1$, or $d_\mathrm{NGC6744} \,=\, 9.6^{+0.5}_{-0.4}$\,Mpc. This is in good agreement with the TRGB distance \citet{Anand2021} measured for NGC\,6744 using deep HST observations, $d_\mathrm{NGC6744} = (9.4 \pm 0.4)$\,Mpc, and within the errors of the \citet{Sabbi2018} measurement, $d_\mathrm{NGC6744} = (8.8 \pm 0.8)$\,Mpc.

This system of galaxies presents several challenges that were not present in the other environments. NGC\,6744 is close enough that all star clusters (which appear as extended sources at this resolution) and potential background contaminants can be identified and masked. Hence, the main concern is not to constrain $P_{\rm r}$, which is negligible, but rather to avoid masking of stellar fluctuations as real sources. This was achieved by adopting a large $\kappa$ value (see Sect. \ref{sec:spurious}). Furthermore, two of the dwarfs are in close proximity to a spiral arm, increasing the possibility of contamination from the host and overlapping dust lanes that could affect the SBF signal; for these galaxies we restricted our measurements to the central regions of the dwarfs, where the signal to noise of the dwarf is highest. Finally, NGC\,6744 covers enough of the field that it likely impacts the background estimate around each dwarf. Thus, our measurements should be treated with caution. Nevertheless, they are consistent with a satellite population around NGC\,6744.

\citep{Carlsten2022} also measured an SBF distance to a sixth dwarf candidate in the image, dw1912m6351 or LEDA\,2815833, placing it behind the NGC\,6744 system at a distance $d=10.62$\,Mpc (note: they obtained systematically lower distances to the other dwarf satellites, making this a background galaxy). We also confirm this as a background galaxy. However, the distance we estimate, roughly the distance of the Fornax cluster, is inherently uncertain; the galaxy exhibits morphological irregularities and appears to have regions of ongoing star formation, meaning that we cannot determine if the signal we measure originates from stellar count fluctuations or another source. Thus, we do not report an exact distance for this galaxy.

\subsection{\label{sc:sspmodels}Comparison with simple stellar population models}

Stellar population models are useful tools to obtain an independent calibration of SBF magnitudes. For this first analysis of SBF in \Euclid data, however, we compare our measurements against simple stellar population (SSP) models to validate the empirical calibration of the $\overline{M}_\mathrm{IE}$ versus colour relation, derived above, and to probe the underlying, unresolved stellar populations of the host galaxies \citep[e.g.][]{Blakeslee2009b,Cantiello2007}. Although composite stellar populations (CSP) are more representative of the mixed populations found in galaxies, and it would be interesting to predict SBF magnitudes with full chemo-evolutionary models, realistic stellar populations may be approximated by linear combinations of SSP models that would simply interpolate within the parameter space \citep[e.g.][]{Blakeslee2001}.

In the following comparisons, we will use the simulated values from the Stellar Population Tools (SPoT) models \citep{Raimondo2005}. The SPoT SSP models for the \Euclid bands are a preliminary release of the updated SPoT code; a complete set of models will be published in a future dedicated paper. In brief, however, the present version of the code generates integrated and SBF magnitudes using the BaSTI (Bag of Stellar Tracks and Isochrones) database of stellar evolutionary tracks \citep{pietrinferni2021,hidalgo2018}, and  derives stellar bolometric corrections in various bands using stellar spectra libraries obtained by merging atmospheric models by \citet{Castelli2003}\footnote{https://wwwuser.oats.inaf.it/fiorella.castelli/} with those published in the MARCS\footnote{https://marcs.astro.uu.se/} database \citep{Gustafsson2008}. 

\begin{figure*}[tbp!]
  \begin{center}
      \includegraphics[width=1.985\columnwidth]{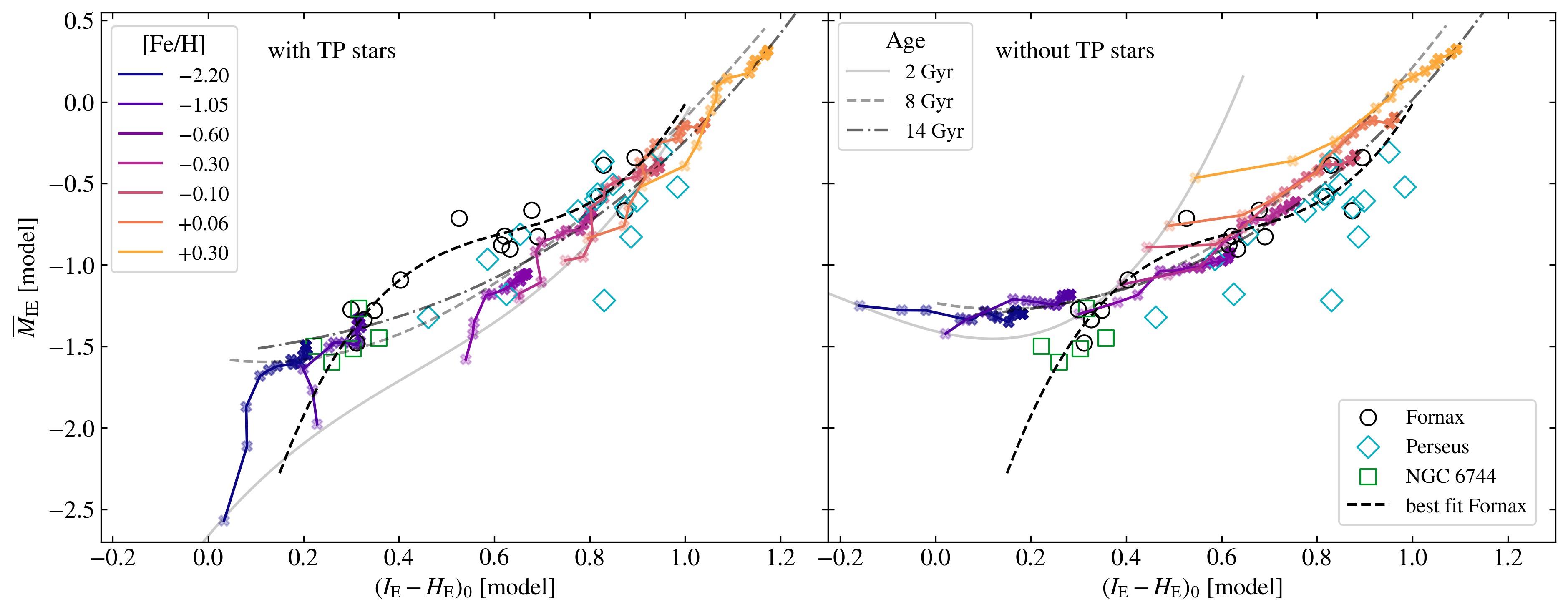}
   \end{center}
\caption{Comparison of the measured SBF values against SPoT SSP models, when thermally pulsating stars are included (\emph{left}) or not (\emph{right}). The $\overline{m}_\mathrm{IE,0}$ measurements in the three groups (Fornax, Perseus, NGC\,6744) have been shifted to a standardized zero point, and converted to $\overline{M}_\mathrm{IE}$ using the distance modulus of Fornax. The SPoT models directly output $\overline{M}_\mathrm{IE}$ and colours, which have been converted to AB magnitudes. Single-value metallicity tracks are distinct colours, with the opaqueness of the points scaling with increasing age. The grey lines are cubic best fits to populations at fixed ages (2\,Gyr, 8\,Gyr, and 14\,Gyr) to help guide the eye.}
\label{fig:SPoT}
\end{figure*}

We directly compare our SBF measurements against the SPoT models in Fig.\,\ref{fig:SPoT}. The SSP models are computed for solar-scaled chemical compositions, with metallicities in the range  $-2.20 \leq \mathrm{[Fe/H]} \leq 0.30$ and ages from $2\,\mathrm{Gyr}$ to $14\,\mathrm{Gyr}$, assuming a total mass $M\sim10^7 M_{\odot}$. The Vega magnitudes output by SPoT have been converted to the AB system, using $m_{\rm AB}-m_{\rm Vega} = -0.268$\, in \IE and $m_{\rm AB}-m_{\rm Vega} = -1.5$ in \HE. We also shifted the galaxies from the Fornax, Perseus, and NGC\,6744 ERO images according to their measured distances, effectively normalizing the $\overline{m}_\mathrm{IE}$ measurements to absolute values. This ensures that all data are anchored to the TRGB distance of Fornax, and that the shape of the stellar population dependence is tied to the calibration for the Fornax galaxies.

For illustrative purposes, we also present two slightly different versions of the SPoT models in Fig.\,\ref{fig:SPoT}. In the left panel, the models include simulated stars in the thermal pulsing asymptotic giant branch (TP-AGB) evolutionary phase, adopting the prescriptions in \citet{Raimondo2005}. The right panel uses the same model parameters, but without any TP-AGB stars.

Although preliminary, both sets of models broadly agree with the observed $\overline{M}_\mathrm{IE,0}$ versus $(\IE{-}\HE)_0$ measurements of our galaxies. We first note that the absolute SBF magnitudes that we measure for the galaxies are of the same order of magnitude as the independently derived models values, suggesting that there are no large systematic errors in our measurements. Furthermore, in the models, the older and more metal-rich stellar populations generate fainter SBF magnitudes and have redder colours than younger, metal-poor systems. This is similar to our measurements, where the most massive galaxies in our sample have redder colours and fainter SBF magnitudes than the dwarfs, which are bluer and brighter. Literature values of the mean stellar ages and total metallicities of the massive galaxies in our Fornax cluster sample confirm that they host old (9.8--13.5\,Gyr), relatively metal-rich ($-0.22\leq [\mathrm{M/H}] \leq 0.21 $) stellar populations in the inner $0.5\,R_\mathrm{e}$ regions \citep{Iodice2019}; these numbers should not be directly compared against the predictions from SSP models, but can be taken as further confirmation of the trends we find. Finally, for a fixed age, the modelled $\overline{M}_\mathrm{IE}$ versus $(\IE{-}\HE)$ relation flattens at bluer colours. We highlight this trend in Fig.\,\ref{fig:SPoT} by fitting a third-order polynomial to the models at three fixed ages (2\,Gyr, 8\,Gyr, and 14\,Gyr) to help guide the eye. Our data follows this trend for colours redder than $(\IE - \HE)_0 \sim 0.4$, but the slope steepens again for the bluest dwarfs; this is largely driven by poor statistics, and more data is needed to determine if this is real. 

Comparing the two sets of models against the measured values in more detail, we note some  differences. The models that include TP-AGB stars have a generally good agreement for redder colours, namely for massive targets whose light is dominated by the emission of  old and metal rich stellar system, and the bluest dwarfs. However, the predicted SBF magnitude is brighter than our measured values at intermediate colours, which we expect to be populated by dwarf galaxies. This could provide some hints about different TP-AGB populations in different galaxy mass regimes, or age/metallicity differences in the dominant stellar components. 

\begin{figure}[tbp!]
  \begin{center}
    \includegraphics[width=0.985\columnwidth]{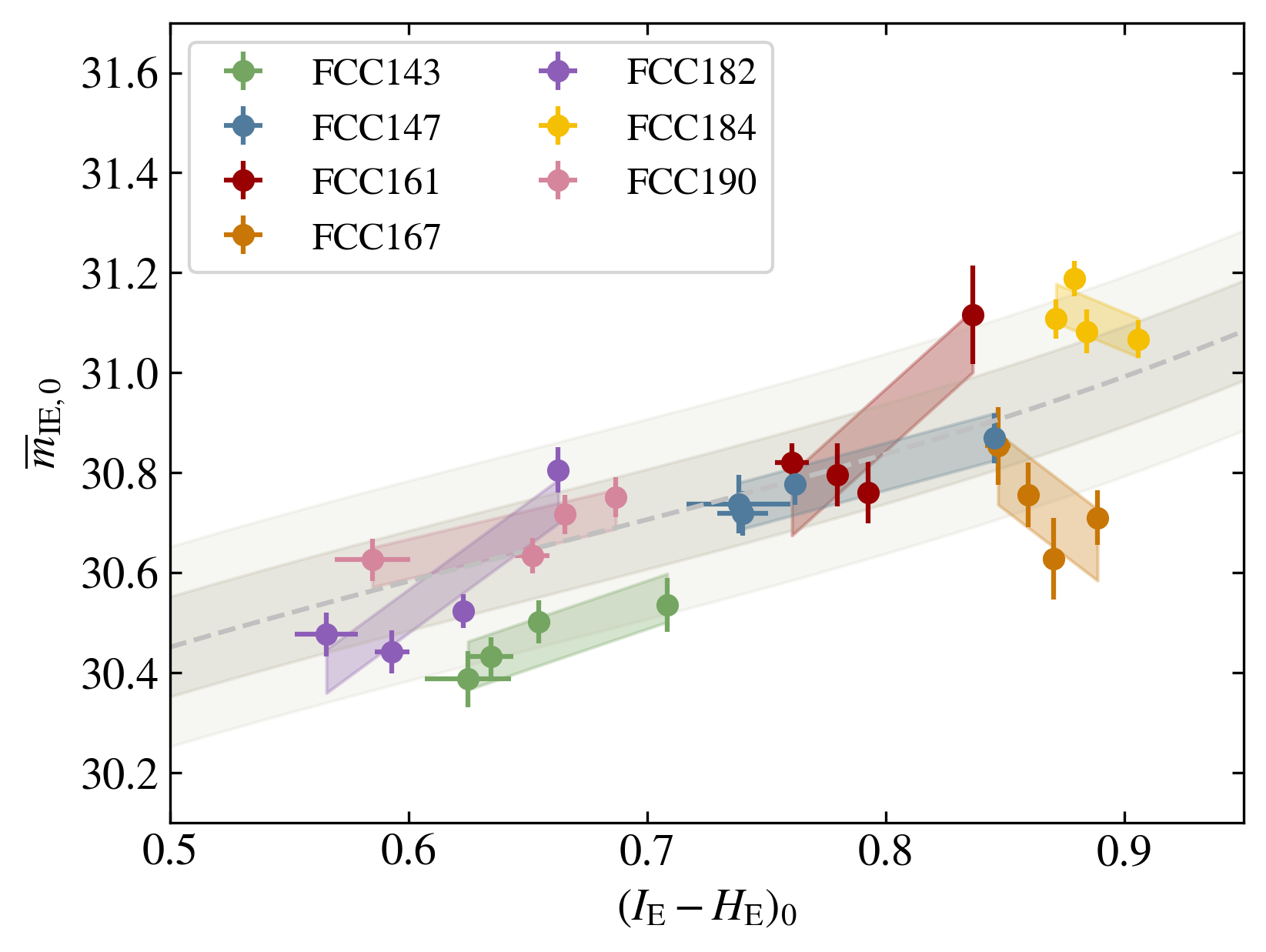}
   \end{center}
\caption{SBF measurements in concentric, independent  annuli can be used to study variations in the underlying stellar populations. Here we measure $\overline{m}_\mathrm{IE,0}$ in four evenly-spaced annuli extending out to roughly $R_\mathrm{e}$ for seven massive galaxies in the sample. For each galaxy, we also highlight the general trend as a shaded region the same colour as the data points; the slope is calculated from linear best fit to the four measurements, while the thickness represents the median error in $\overline{m}_\mathrm{IE,0}$. The galaxies are superimposed on the best-fit line for the entire sample (grey dashed line), and the grey shaded region corresponds to $\overline{m}_\mathrm{IE,0} \pm 0.1$.}
\label{fig:pop_gradients}
\end{figure}

\begin{figure*}[tbp!]
  \begin{center}
      \includegraphics[width=1.985\columnwidth]{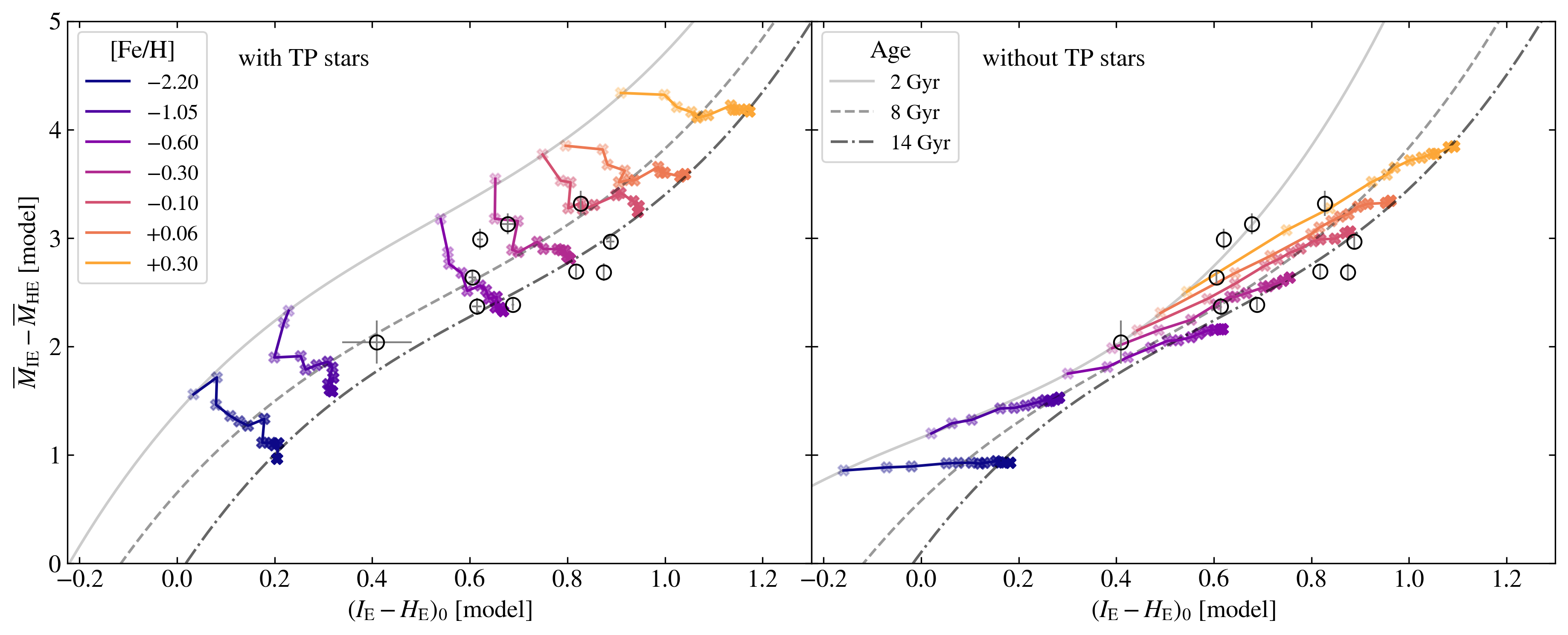}
   \end{center}
\caption{Preliminary $\overline{M}_\mathrm{IE} - \overline{M}_\mathrm{HE}$ measurements compared with SSP models. The data points (open black circles) should be treated with caution (see discussion in the text) and the error bars on the SBF colour, dominated by the errors from the \HE measurements, represent lower limits. \emph{Left panel}: models including thermally pulsating stars. \emph{Right panel}: models excluding thermally pulsating (TP) stars. Colours and symbols for the models are the same as in Fig.\,\ref{fig:SPoT}.
}
\label{fig:SPoT_IEvsHE}
\end{figure*}

The agreement between our measured values and the models without any TP-AGB stars appears to be excellent upon first inspection. As seen in the right panel of Fig.\,\ref{fig:SPoT}, the differences between our calibration line (Eq.\,\ref{eq:fit}) and the third-order best fit to the 14\,Gyr population is within the calibration errors of both the models and our empirical relation. 
However, we also observe that the position of the massive galaxies along the TP-AGB-free SSP relation would indicate that they are dominated by stars with sub-solar metallicities, which is unexpected \citep[e.g.,][]{Iodice2019}. In the models that include TP-AGB stars, the mean metallicity is closer to the solar value, more in line with expectations from independent studies.

The TP-AGB phase experienced by low- and intermediate-mass stars is one of the most uncertain phases of stellar evolution as many physical processes are involved in the modelling of these stars (e.g., envelope convection, chemical mixing, mass loss, stellar wind, rotation, and magnetic fields) and the models need to be calibrated with the aid of observations \citep{wagenhuber1998}. Although preliminary, our comparison may indicate that the contribution of TP-AGB stars depends on the age and metallicity of the dominant stellar population, corroborating previous results from \citet{Cantiello2003}. This dependence could be the result of more efficient mass-loss rates in low-metallicity stars \citep[see, e.g.,][]{Girardi2010,vanLoon2006}, which would then lead to a lower number of TP-AGB stars in metal-poor stellar populations and fainter SBF magnitudes.

In Fig.\,\ref{fig:SPoT}, there are some outliers in our measured values that deviate from the parameter space of the models. These galaxies are predominantly from the Perseus cluster sample, where the acceptable (but not optimal) measurements have a higher scatter. The scatter is also driven by distance effects; our measurements indicate that 2MASX\,J03171358+4126082 is in the foreground of the Perseus cluster, but all the galaxies in the Perseus sample have been shifted by the relative distance of the cluster for this test. 

Work is in progress to optimize the models, including the contribution from TP-AGB stars to better improve the  agreement between the data and models over the entire colour, metallicity, and mass range that we probe.  Nevertheless, the agreement between these two completely independent approaches supports the reliability of the measurements presented here, even though further refinements are still required.

Previous work has also measured SBF gradients in individual galaxies \citep[e.g.][]{Cantiello2007}. As discussed by \citet{Cantiello2011,Cantiello2005}, these gradients trace population variations in the unresolved stellar populations, which can be linked to different formation scenarios. In Fig.\ref{fig:pop_gradients}, we show the $\overline{m}_\mathrm{IE,0}$ versus $(\IE{-}\HE)_0$ measurements in concentric, independent annuli for seven massive galaxies in the sample. FCC\,148 and FCC\,170 were dropped from this analysis due to the proximity of neighboring galaxies (FCC\,148) or the complexity of internal structures (FCC\,170) that reduced the region in which reliable SBF measurements can be made. For the remaining galaxies, we probed from the centre-most regions ($R_\mathrm{in}$ in Table\,\ref{table:results}) to approximately $R_\mathrm{e}$, splitting the region into four, evenly-spaced annuli. The central regions of each galaxy in the sample are redder than the outskirts, allowing one to easily trace the radial trends in Fig.\,\ref{fig:pop_gradients}. We observe a range of gradients, with some (e.g., FCC\,190 and FCC\,147) following the best fit magnitude-colour relation almost perfectly, while others (e.g., FCC\,167) are much steeper. A more detailed study of these gradients will be the subject of future work.

Given the uncertainties in the models (e.g., the contributions from TP-AGB stars and the overlap between metallicity tracks and the fixed age best-fit lines), it is not clear if these slopes should be interpreted as metallicity-dominated or age-dominated radial changes. However, the ability to measure SBF gradients with VIS data, demonstrated here, combined with the larger sample of measurements expected from upcoming \Euclid releases, will allow us to better study the underlying physical processes driving the observed relations.

\subsection{\HE SBF measurements: colours versus model predictions}
As discussed in the introduction, NIR bands are generally preferred for SBF analysis, however we have worked with the \IE band due to known limitations of \Euclid's NISP imaging for SBF (in particular, the use of a bilinear kernel and an undersampled PSF). Nevertheless, we tested the potential to measure SBF magnitudes from the \HE images. At this stage, given the preliminary nature of the measurements, we focus on the ten brightest Fornax cluster galaxies and focus on analysing the SBF colors ($\overline{M}_\mathrm{IE} - \overline{M}_\mathrm{HE}$).

Beyond their use in distance studies, the combination of \IE and \HE SBF measurements provides a tool to study stellar population properties in the target galaxies, independently of classical photometry and/or spectroscopy. The results of the combined $\overline{M}_\mathrm{IE}$ and $\overline{M}_\mathrm{HE}$ measurements versus the classical integrated colours $(\IE{-}\HE)$ of the target galaxies are shown in Fig.\,\ref{fig:SPoT_IEvsHE}. The plot shows the models with TP-AGB stars in the left panel and without TP-AGB stars in the right panel. 

Despite the uncertainties in the measurement of $\overline{m}_\mathrm{HE}$, the comparison between models and observations show good agreement, providing a further indication of the reliability of both. Unlike the comparison between the models and the \IE measurements, however, here it appears that the models with TP-AGB stars are more consistent with the data than the models without TP-AGB stars. The SSP tracks without TP-AGB stars would suggest that several galaxies are dominated by relatively young stellar populations with ages younger than 10\,Gyr and super-solar metallicities. This is also inconsistent with measurements from the literature \citep{MartinNavarro2019,Iodice2019,Rakos2001}. In contrast, the models with TP-AGB stars appear to be in agreement with the expected age and metallicity ranges for these galaxies. Of course, these results should not be over-interpreted as we are still optimizing the data reduction and the measurement pipeline for the NISP passbands. The values are reported here to show how the combination of SBF colour measurements, integrated colours, and stellar population models represent an effective method for studying unresolved extragalactic stellar populations or, conversely, for refining our understanding of stellar evolution scenarios \citep{Blakeslee2001,Cantiello2011,Jensen2015}. This is especially relevant for the evolutionary phases of low-mass, luminous stars which have the strongest impact on the SBF signal but are statistically rare, making them difficult to study using classical stellar photometry methods \citep{Raimondo2009,Kim2021,RodriguezBeltran2024}.

\section{\label{sc:conclusions}Summary and conclusions}

SBF is a powerful technique to measure distances to nearby galaxies from photometry alone. Large surveys such as \Euclid are expected to observe thousands of galaxies for which an SBF signal can be detected, potentially allowing us to map the nearby Universe with unprecedented accuracy. However, current implementations of the SBF method typically require a significant amount of human intervention to obtain the final measurements. In this work, we present the first application of the \texttt{python} code {\texttt{Flexible Automated Self-contained Tool for SBF (FAST-SBF)}} to \Euclid data. 

The image quality of \Euclid is clearly sufficient to detect the SBF signal in the images selected for this analysis, although the standard data reduction pipeline is not ideal for fluctuation measurements. Nevertheless, we demonstrated that reliable $\overline{m}$ can be recovered with the ERO data reduction strategy, where a Lanczos3 kernel has been adopted for the image stacking. Future data products from the Euclid MERge Processing Function (MER PF; \citealt{Q1-TP006}) data reduction pipeline are expected to use bilinear interpolation -- ideal for the cosmological science goals of \Euclid, but not for SBF measurements -- for all filters, thus alternate data reduction strategies will need to be explored going forward. Teams are already working to provide the community with such data products (e.g. Dimauro et al., in prep; Bamford et al., in prep). With these images, it may also be possible to measure the SBF signal in one of the NIR bands, where the stellar fluctuation amplitudes are brighter and less affected by contaminating dust.  

We first tested the procedure and our {\texttt{FAST-SBF}} code on galaxies in the well-studied Fornax galaxy cluster, whose distance has been constrained by prior redshift-independent distance measurements, including SBF. The massive and dwarf galaxies -- at least, those that are large enough to obtain good statistics and not contaminated by light from the neighbouring galaxies -- follow a tight $\overline{m}_\mathrm{IE,0}$ versus $(\IE{-}\HE)_0$ correlation. 

We calibrated the best-fit line to the Fornax data, to provide a zero point for the method. This was then tested on galaxies in the Perseus cluster and the dwarf satellites around NGC\,6744. Assuming the distance to the Fornax cluster is 19.3\,Mpc, we estimate a distance of $d=69.5$\,Mpc for the Perseus cluster and $d = 9.6$\,Mpc for NGC\,6744, in good agreement with previous measurements. In Perseus, we identify one galaxy as a foreground object; it has a velocity that is consistent with the cluster, but this can be reconciled with our measurements if it is a foreground galaxy moving towards the cluster at a high velocity. This highlights the capabilities of SBF in \Euclid data, and to map galaxies in the nearby Universe.

To further validate the measurements and the calibration approach adopted, we also compared our results with SSP model predictions. Using the SPoT models with preliminary runs in the \Euclid passbands, we find good agreement between the data and models. The absolute SBF magnitudes returned from these two independent estimates are of the same order of magnitude, suggesting that there are no large systematic errors in our results. The general trend from the models -- a steep slope for redder (more metal-rich) galaxies that flattens for bluer (more metal-poor) systems -- largely mirrors our observations for both massive and dwarf galaxies. In the specific case of the SSP models without TP-AGB stars, we find a striking match between the models and the empirical relation derived from the Fornax data.

A key limitation in our current Fornax-based calibration, however, is the small sample. The current calibration is highly sensitive to the bluest galaxies used for the fit and may be biased by the fact that dwarf cluster members are not expected to be particularly metal-poor. The cluster will be re-imaged as part of the EWS, which will provide enough galaxies to firmly establish the $\overline{M}_\mathrm{IE}$ versus $(\IE - \HE)_0$ relation across a large range of masses, probe the scatter in the relation, and test the limitations of the method. Moreover, these results, combined with the measurement of SBF amplitudes from NISP data -- where the contribution of red and cold TP-AGB stars is expected to be more significant -- demonstrate that SBF magnitudes and colours can be efficiently used to refine stellar population models, allowing us to better constrain the properties of stars associated with the fastest and brightest evolutionary stages in old stellar systems.

The EWS is expected to observe thousands of dwarf galaxies suitable for SBF analysis (e.g., dwarf ellipticals) in the local Universe \citep{Q1-SP001}. To fully benefit for this vast new dataset, it will be crucial to revisit the calibration of SBF magnitudes for extremely blue, metal-poor systems. The SPoT models suggest the calibration should remain relatively flat for fixed-age galaxies with colours $(\IE{-}\HE)_0 < 0.5$. This expectation can be directly tested with upcoming \Euclid data releases.

Future \Euclid observations will enable a significant expansion of the results in this work. The wide coverage will provide an unprecedented number of galaxies for SBF measurements, allowing us to determine cosmic distances and study stellar populations with far greater statistical power.

\begin{acknowledgements}
We are grateful to Adriano Pietrinferni and Santi Cassisi for providing us tables with key points of their evolutionary tracks published in Hidalgo et al. (2018), and Pietrinferni et al. (2021). R.H.  and M.C. acknowledge support from the project ``INAF-EDGE'' (Large Grant 12-2022, P.I. L.\,Hunt). Part of this work was supported through the INAF “Astrofisica Fondamentale” GO-grant 2024 n. 12 (PI: M. Cantiello). MC acknowledges support from the ASI–INAF agreement “Scientific Activity for the Euclid Mission” (n. 2024-10-HH.0; WP8420) and from the ESO Scientific Visitor Programme. N.H. acknowledges support from the Polish National Science Centre grant 2023/50/A/ST9/00579. M.P. is supported by the Academy of Finland grant No. 347089.
%ADD ACKNOWLEDGEMENT TO ASTROPY, SCIPY, NASA ADS, IRSA DUST, DS9.  
  \AckERO 
  \AckEC 
\end{acknowledgements}

\bibliography{Euclid, Q1, sbf}

\begin{appendix}

\section{\label{appendix:fits}Output of FAST-SBF for the galaxies in the Fornax cluster with prior SBF measurements}
Similar to Fig.\,\ref{fig:process_example1}, we include image cutouts, the model residuals, the masked aperture used for the SBF analysis, and the azimuthally averaged power spectrum for a selection of the other massive galaxies in Fornax. Remarks on individual galaxies are included in Appendix\,\ref{appendix:individual}.

\begin{figure*}[htbp!]
  \begin{center}
    \includegraphics[width=1.98\columnwidth]{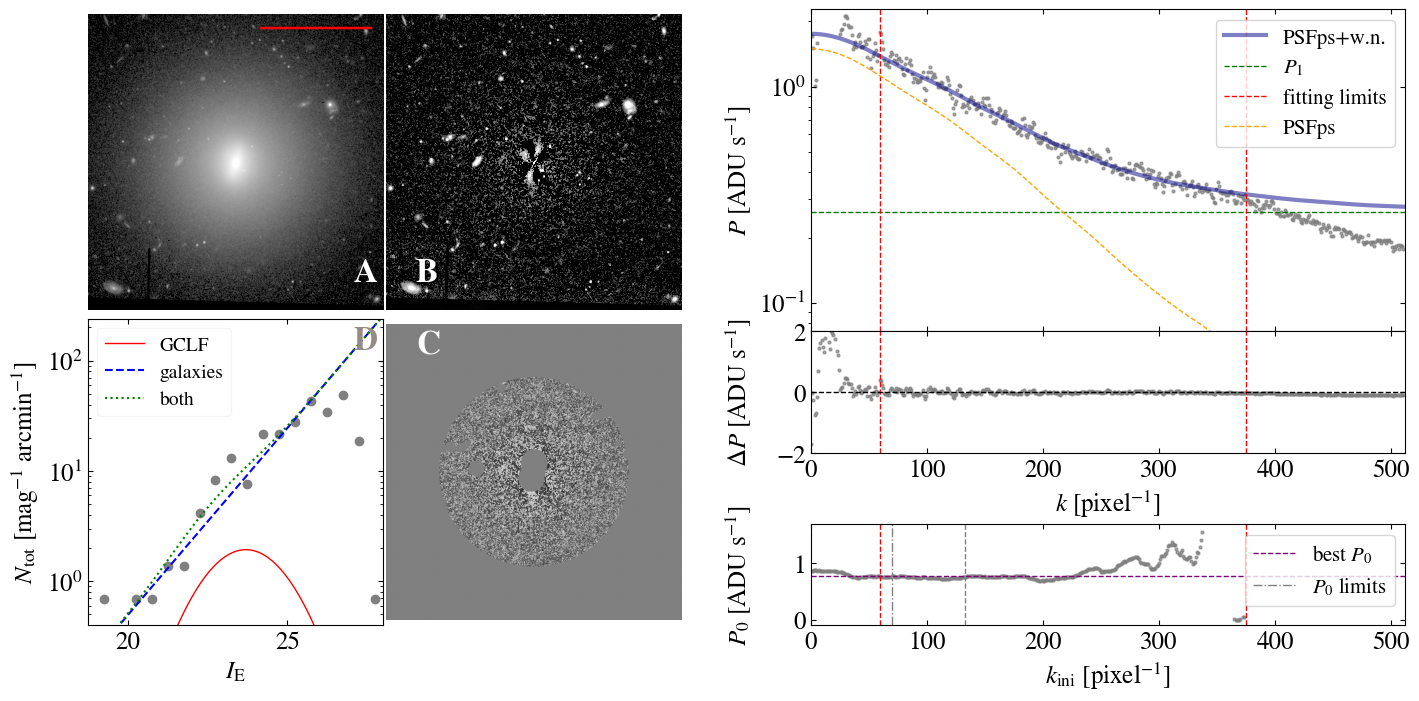}
   \end{center}
\caption{Same as Fig.\,\ref{fig:process_example1}, but for FCC\,182.}
\label{fig:example2}
\end{figure*}

\begin{figure*}[htbp!]
  \begin{center}
    \includegraphics[width=1.98\columnwidth]{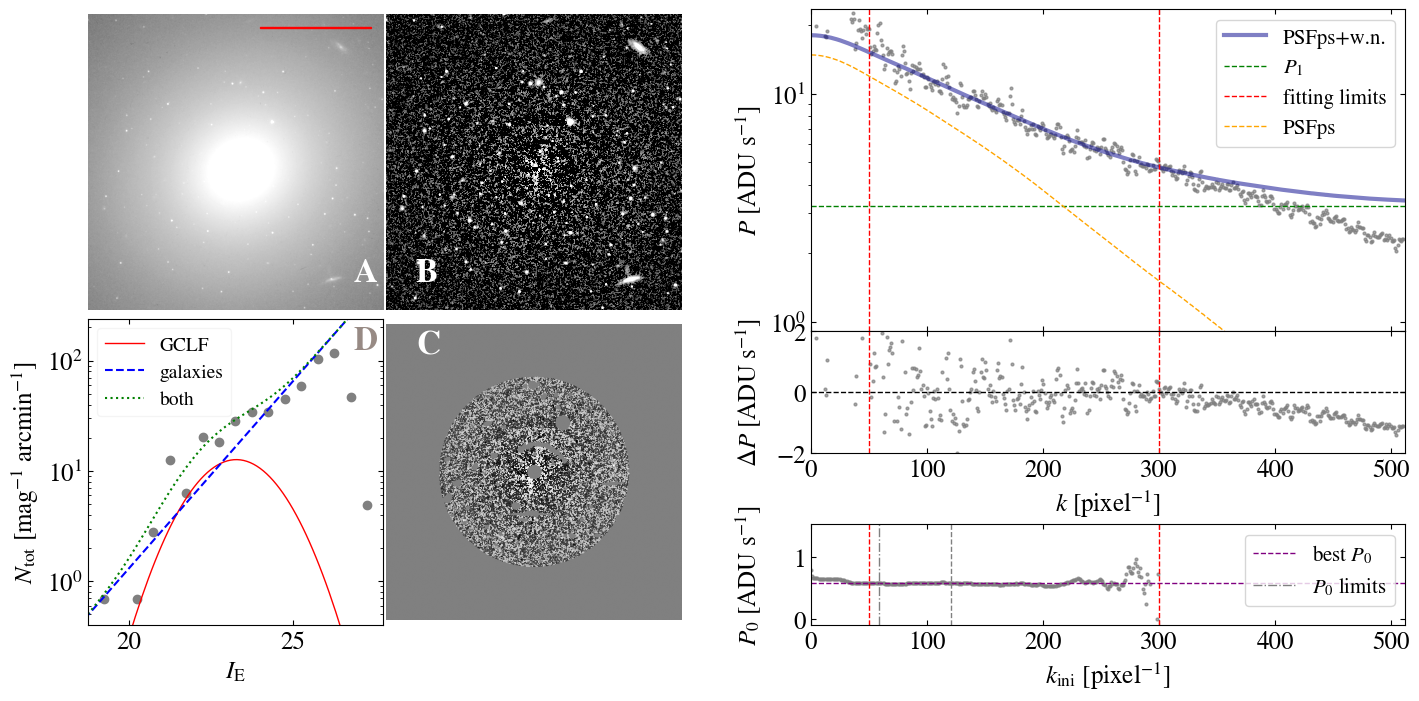}
   \end{center}
\caption{Same as Fig.\,\ref{fig:process_example1}, but for FCC\,147.}
\label{fig:example3}
\end{figure*}

\begin{figure*}[htbp!]
  \begin{center}
   \includegraphics[width=1.98\columnwidth]{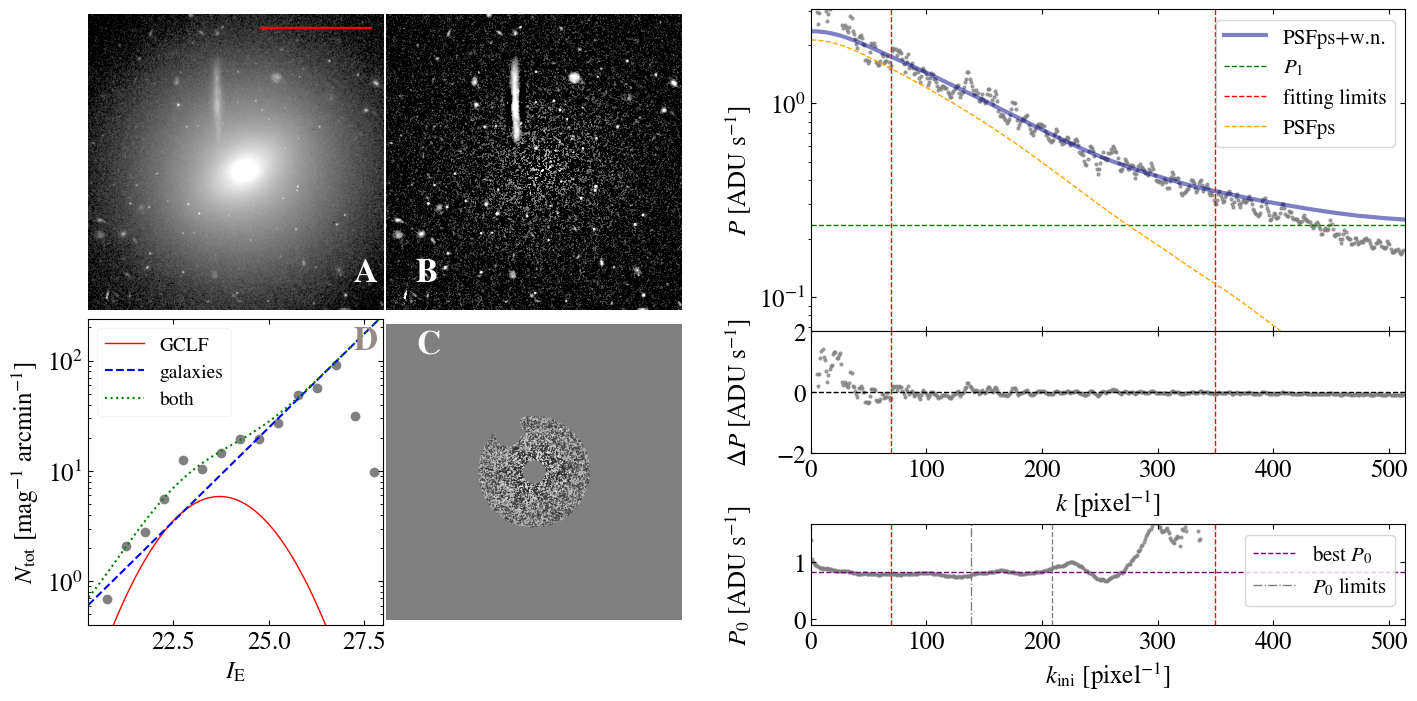}
   \end{center}
\caption{Same as Fig.\,\ref{fig:process_example1}, but for FCC\,143.}
\label{fig:example4}
\end{figure*}

\section{\label{appendix:individual}Comments on the fits of individual galaxies}

Several galaxies in the sample present unique features that impact the galaxy modelling. Below are notes on individual galaxies in the Fornax cluster (Appendix\,\ref{notes:fornax}) and the NGC\,6744 system (Appendix\,\ref{notes:ngc6744}). For this first analysis, galaxies in the Perseus cluster with visible substructures or evidence of recent star formation were removed from the sample, and therefore we do not provide commentary on any galaxies in that cluster. 

\subsection{Fornax galaxy cluster sample}
\label{notes:fornax}

\noindent FCC\,143 \hfill \newline
The inner isophotes of FCC\,143 are more elongated than the outer isophotes, which leads to some large-scale structures in the model residual. There is also an edge-on galaxy that overlaps due to projection effects, but it is easy to mask. \citet{Iodice2019} find evidence for an ongoing interaction between FCC\,143 and FCC\,147 with SBF distance estimates from \citet{Blakeslee2009} putting these galaxies basically at the same projected distance, within uncertainties. They place FCC\,143 slightly in the foreground of FCC\,147, which is also supported by our measurements. 
 \\

\noindent FCC\,147 \hfill \newline
FCC\,147 is a massive elliptical with fairly regular isophotes and minimal interference from regions with no data. It is close in physical space to both FCC\,143 and FCC\,148, with the latter in the outer halo of the galaxy. However, we do not measure out to those distances, so we do not expect any contamination from the neighbouring galaxies. \\

\noindent FCC\,148 \hfill \newline
Although FCC\,148 is in the halo of its massive companion, FCC\,147, we  expect minimal contamination from the neighbouring galaxy, given the relative brightness of the two galaxies at the centre of FCC\,148. We restrict the measurements to the inner portions of the galaxy, but cannot use the very centre as there is a region of pixels with no data going through the middle of the galaxy. If we model FCC\,148 on the original image and an image from which the model of FCC\,147 has been subtracted, we obtain similar $\overline{m}$ values.  \\

\noindent FCC\,167 \hfill \newline
This galaxy has a highly inclined dusty disk at the centre of the galaxy, which makes the inner 6$''$ unsuitable for SBF measurements. The modelling of this galaxy is also complicated by a region of pixels with missing data that intersects the galaxy approximately $1'$ from the centre, and a diffuse excess of light at the other end of the galaxy. \citet{MartinNavarro2019} used MUSE data to study age and metallicity gradients in this galaxy and found a strong radial metallicity trend. \\

\noindent FCC\,170 \hfill \newline
This galaxy is alternatively classified as a lenticular galaxy or an edge-on spiral. It exhibits a narrow disk and a clear, boxy bulge. For this galaxy, we masked the disk before modelling the galaxy, and excluded it from the SBF measurements. Of all the galaxies, this one shows the widest colour differences in different annuli. \citet{Pinna2019} used MUSE data to study the stellar populations in the galaxy. It has been noted that it is more difficult to measure SBF in edge-on spirals \citep{Ciardullo2012,Mei2007}, and indeed \citet{Spriggs2021} find the largest disagreements between SBF and PNLF distance estimates for the edge-on spirals.  \\

\noindent FCC\,182 \hfill \newline
Compared to the other massive galaxies, FCC\,182 is relatively small and isolated, with a region of missing data values only affecting the outskirts of the galaxy. However, the morphology of the galaxy is not smooth, with twisted isophotes as one moves to the outskirts. This results in large-scale structures in the residual image. We therefore avoid the centre when measuring the SBF signal. 
\\

\noindent FCC\,184 \hfill \newline
The modelling of this galaxy is affected by the presence of a compact, nearly face-on disk at the centre ($r \approx 10''$), a diffuse bar-like structure that extends out to $r\approx 25''$, and a region of missing data that bisects the galaxy. \\

\noindent FCC\,190 \hfill \newline
The isophotes in FCC\,190 are twisted, making this galaxy difficult to model. It appears to have a double bar, aligned more or less perpendicularly to each other, surrounded by a regular halo. \\

\subsubsection{Dropped dwarf candidates}
As noted in the Sect.\,\ref{sc:calibration}, several dwarf candidates were dropped from the analysis because of their proximity to a massive galaxy in the image. These include: LEDA\,74742, LEDA\,74747, LEDA\,74761, LEDA\,74754, LEDA\,74757, and LEDA\,74796. As objects in the HyperLeda tables, these are relatively bright and extended dwarf galaxies, and many have fluctuations in the surface brightness that can be seen by eye. These galaxies either have SBF magnitudes that are too bright for their colour, or are much redder for the measured $\overline{m}_0$ than one would expect for a member of the Fornax cluster. We do not believe them to be foreground objects, and instead attribute this to contamination from neighbouring galaxies (see Fig.\,\ref{fig:sky} for a measurement of the SBF signal in the halo of FCC\, 147, at the approximate radial distance of LEDA\, 74747).

\subsection{NGC 6744 dwarf satellites}
\label{notes:ngc6744}
\noindent dw1908m6343 \hfill \newline
This is an elongated dwarf galaxy that partially overlaps with a spiral arm of NGC\,6744. \citet{Carlsten2022} included it as a satellite of NGC\,6744 based on the visual appearance of the fluctuations, even though the distance they measured was too small; this was attributed to the fact that it is a very blue, star forming irregular dwarf. In the \IE imaging, one can clearly see an increase in bright, compact sources away from the centre of this cigar shaped galaxy. We restricted the SBF annulus to the central region of the dwarf for this analysis, and measure a colour that is consistent with other dwarfs in the image.\\

\noindent dw1911m6413 \hfill \newline
This is a large dwarf galaxy with a slightly asymmetric light profile, possibly due to irregular patches of intermediate/young stellar populations. Nevertheless, the model residual, after masking the irregular regions, appeared clean enough to measure the amplitude of the SBF signal. This is the dwarf galaxy in the NGC\,6744 field that has a redshift measurement \citep{Tully2016}. \\

\noindent dw1906m6357 \hfill \newline
This is a relatively small, nucleated dwarf elliptical galaxy that is adjacent to a spiral arm. Conveniently, there appears to be a gap in the spiral arm at this position, which allows us to obtain cleaner measurements. While dust in the spiral arm could dampen the SBF signal, we do not believe that is an issue here, and we obtain a measurement that is consistent with the other satellite dwarfs.

\section{\label{appendix:tables}Data tables}
Here we report measurements for the candidate cluster members. As discussed in Sect.\,\ref{sc:calibration} (and see the notes in Appendix\,\ref{appendix:individual}) these are galaxies with a measurable SBF signal that, given the relative isolation of the Fornax cluster, suggests they are cluster members even though some aspect of the output numbers are questionable. In Table\,\ref{table:maybe}, we include basic properties of these galaxies.

\begin{table*}[tbp!]
\newcommand{\pd}{\phantom{1}}
\setlength{\tabcolsep}{7.25pt}
\caption{Dwarf galaxies that we tentatively classify as Fornax cluster members; see Sect.\,\ref{sc:calibration} for a discussion. The last two entries, with galaxy names that begin with ``g", have no known matches in the literature. The coordinates are from the catalogues output by {\texttt{JAFAR}} during the original creation of the sample. We report only the effective radii, \IE of the galaxy, measured colours -- which are not extinction corrected, and the extinction correction.    }
\smallskip
\label{table:maybe}
\smallskip
\centering
\small
\begin{tabular}{l l l l r l l c }
  \hline
  & & & & & &\\[-7pt]
\omit\hfil Name\hfil & Alt Name & \omit\hfil RA \hfil & \omit\hfil Dec\hfil  & \omit\hfil $R_\mathrm{e}$ \hfil  &\omit\hfil \IE \hfil &\omit\hfil $(\IE{-}\HE)$\hfil &  \omit\hfil E(B-V) \hfil \\
& & \omit\hfil[deg] \hfil & \omit\hfil\,[deg] \hfil & [arcsec]  & & &\\
  & & & &  & & &\\[-8pt]
\hline
  & & & & & & & \\[-8pt]
FCC\,171 & $\cdots$ & 54.155270 & $-$35.385904  & 12.64 &  17.52 &$\phantom{-}$0.613$\pm$0.053 & 0.0109 \\ 
FDS\,110299 & $\cdots$ & 54.410736 & $-$35.385691  & 5.71 &  19.47 &$\phantom{-}$0.188$\pm$0.115 & 0.0128 \\ 
FDS\,110306 & $\cdots$ & 54.249219 & $-$35.343329  & 6.08 &  18.91 &$\phantom{-}$0.153$\pm$0.079 & 0.0116 \\ 
FDS\,160141 & $\cdots$ & 53.716940 & $-$35.623429  & 2.10 &  20.46 &$\phantom{-}$0.305$\pm$0.060 & 0.0127 \\ 
FDS\,160230b & $\cdots$ & 53.800059 & $-$35.434134  & 1.70 &  22.16 &$\phantom{-}$0.384$\pm$0.186 & 0.0124 \\ 
FCC\,140 & LEDA\,74742 & 53.735340 & $-$35.190876  & 7.13 &  18.27 &$\phantom{-}$0.447$\pm$0.039 & 0.0119 \\ 
 FCC\,144 & LEDA\,74745 & 53.750844 & $-$35.322396  & 3.72 &  19.05 &$\phantom{-}$0.137$\pm$0.060 & 0.0119 \\ 
FCC\,145 & LEDA\,74747 & 53.772810 & $-$35.218471  & 5.92 &  18.40 &$\phantom{-}$0.368$\pm$0.092 & 0.0118 \\ 
FCC\,160 & LEDA\,74761 & 54.016930 & $-$35.388845  & 11.64 &  17.04 &$\phantom{-}$0.499$\pm$0.035 & 0.0105 \\ 
FCC\,168 & LEDA\,74770 & 54.116574 & $-$35.210698  & 6.07 &  18.40 &
$\phantom{-}$0.291$\pm$0.087 & 0.0124 \\
 FCC\,185 & LEDA\,74783 & 54.261670 & $-$34.875560  & 3.60 &  19.36 &$\phantom{-}$0.274$\pm$0.052 & 0.0141 \\ 
FCC\,197 & LEDA\,74796 & 54.421380 & $-$35.296254  & 4.51 &  18.67 &$\phantom{-}$0.151$\pm$0.043 & 0.0130 \\ 
FDS\,11\,LSB48 & NGFSJ033602-351839 & 54.007612 & $-$35.310800  & 2.82 &  21.23 &$-$0.012$\pm$0.133 & 0.0111 \\ 
g15 & $\cdots$ & 53.714600 & $-$35.076567  & 2.46 &  22.75 &$\phantom{-}$0.142$\pm$0.389 & 0.0120 \\ 
g89 & $\cdots$ & 54.420153 & $-$34.933085  & 0.89 &  21.19 &$\phantom{-}$0.157$\pm$0.066 & 0.0120 \\
\hline
\end{tabular}
\label{LastPage}
\end{table*}

\end{appendix}

\end{document}